\documentclass[12pt,onecolumn]{IEEEtran} 
\usepackage{amsmath,amsfonts}
\usepackage{algorithmic}
\usepackage{algorithm}
\usepackage{array}
\usepackage{amssymb,amsfonts}
\usepackage{color}
\usepackage{textcomp}
\usepackage{stfloats}
\usepackage{url}
\usepackage{verbatim}
\usepackage{graphicx}
\usepackage{cite}
\usepackage{cases}
\usepackage{setspace} 

\usepackage{caption}
\usepackage{subcaption}

\newtheorem{Th}{Theorem}
\newtheorem{Lm}{Lemma}
\newtheorem{propRemark}{Remark}

\usepackage{textcomp}
\usepackage{xcolor}

\begin{document}
\baselineskip 4.1ex 

\title{Energy-Efficient Beamforming Design for Integrated Sensing and Communications Systems}

\author{Jiaqi Zou,~\IEEEmembership{Graduate Student Member,~IEEE}, Songlin Sun,~\IEEEmembership{Senior Member,~IEEE}, Christos Masouros,~\IEEEmembership{Senior Member,~IEEE}, Yuanhao Cui,~\IEEEmembership{Member,~IEEE}, \\ Ya-Feng Liu,~\IEEEmembership{Senior Member,~IEEE}, and Derrick Wing Kwan Ng,~\IEEEmembership{Fellow,~IEEE}

\thanks{Part of this work has been submitted to the IEEE Global Communications Conference (GLOBECOM 2023) for possible presentation~\cite{zou2023sensing}.}
\thanks{Jiaqi Zou is with the School of Information and Communication Engineering, Beijing University of Posts and Telecommunications (BUPT), Beijing 100876, China, and also with the Department of Electrical and Electronic Engineering, University College London, London WC1E 7JE, UK (e-mail: jqzou@bupt.edu.cn). }
\thanks{Songlin Sun and Yuanhao Cui are with Beijing University of Posts and Telecommunications (BUPT), Beijing, China (e-mail: slsun@bupt.edu.cn, cuiyuanhao@bupt.edu.cn).}
\thanks{Christos Masouros is with the Department of Electrical and Electronic Engineering, University College London, WC1E 7JE, UK (e-mail: chris.masouros@ieee.org). }
\thanks{Ya-Feng Liu is with the State Key Laboratory of Scientific and Engineering Computing, Institute of Computational Mathematics and Scientific/Engineering Computing, Academy of Mathematics and Systems Science, Chinese Academy of Sciences, Beijing 100190, China (e-mail: yafliu@lsec.cc.ac.cn)}
\thanks{Derrick Wing Kwan Ng is with the School of Electrical Engineering and Telecommunications, University of New South Wales, Sydney, NSW 2052, Australia (e-mail: w.k.ng@unsw.edu.au).}
}

\markboth{~}%
{Shell \MakeLowercase{\textit{et al.}}: A Sample Article Using IEEEtran.cls for IEEE Journals}


\maketitle

\begin{abstract}
\begin{spacing}{1.5} 
In this paper, we investigate the design of energy-efficient beamforming for an ISAC system, where the transmitted waveform is optimized for joint multi-user communication and target estimation simultaneously. 
We aim to maximize the system energy efficiency (EE), taking into account the constraints of a maximum transmit power budget, a minimum required signal-to-interference-plus-noise ratio (SINR) for communication, and a maximum tolerable Cram\'{e}r-Rao bound (CRB) for target estimation. 
We first consider communication-centric EE maximization.
To handle the non-convex fractional objective function, we propose an iterative quadratic-transform-Dinkelbach method, where Schur complement and semi-definite relaxation (SDR) techniques are leveraged to solve the subproblem in each iteration. 
For the scenarios where sensing is critical, we propose a novel performance metric for characterizing the sensing-centric EE and optimize the metric adopted in the scenario of sensing a point-like target and an extended target.
To handle the nonconvexity, we employ the successive convex approximation (SCA) technique to develop an efficient  algorithm for approximating the nonconvex problem as a sequence of convex ones. 
Furthermore, we adopt a Pareto optimization mechanism to articulate the tradeoff between the communication-centric EE and sensing-centric EE. We formulate the search of the Pareto boundary as a constrained optimization problem and propose a computationally efficient algorithm to handle it. 
Numerical results validate the effectiveness of our proposed algorithms compared with the baseline schemes and
the obtained approximate Pareto boundary shows that there is a non-trivial tradeoff between communication-centric EE and sensing-centric EE, where the number of communication users and EE requirements have serious effects on the achievable tradeoff.
 \end{spacing} 
\end{abstract}


\section{Introduction}

Integrated sensing and communications (ISAC) are anticipated as a viable enabling technology for unlocking the potential of next-generation wireless networks, as the two kinds of systems tend to share various common devices, signal processing techniques, and even the hardware circuitries. Rather than the conventional parallel development of the two systems, the joint designs advocating their coexistence and cooperation have attracted extensive research interest in recent years. For instance, the coexistence of communication and radar systems focuses on spectrum sharing or physical integration design, which mainly aims to mitigate the mutual interference and efficiently manage the limited wireless resources~\cite{chiriyath2017radar}. Indeed, since communication and radar systems may transmit independent signals superimposed in the time/frequency domains, the interference between each other should be minimized to facilitate their individual functionalities. In such cases, numerous approaches have been proposed, such as cooperative spectrum sharing~\cite{hayvaci2014spectrum} and beamforming design~\cite{huang2015radar}. Nevertheless, the existence of inevitable mutual interference still causes certain limitations on spectral efficiency performance.

Meanwhile, compared with the coexistence design approaches that generate communication and sensing signals separately, ISAC employs a common transmitted signal for realizing communication and sensing simultaneously. In such a case, the crux of ISAC is how to design a specialized waveform for effectively transmitting data and sensing potential targets. 
In particular,  the waveform design can be categorized into the communication-centric, radar-centric, and joint design according to the design goals~\cite{zhang2021enabling,wangchaoqos}. Specifically, the radar-centric design aims to modulate the communication data onto the radar pulses, where the radar probing signals can be regarded as an information carrier~\cite{mealey1963method}. On the other hand, communication-centric approaches utilize existing communication signals to sense the environment, such as cellular signals~\cite{david2015cellular} and Wi-Fi signals~\cite{ma2019wifi,zhangiot9198891}. In particular, various environmental conditions can be extracted from the received echoes of the communication signals, as the target's existence or movement inevitably affects the signal's propagation. Nevertheless, the integration performance is limited in the above two approaches, as the communication/sensing functionality is often carried out as ancillary tasks. In contrast, the joint ISAC design studies the co-design of signaling methodologies enabling both communications and sensing, which is the research content of this work.


\subsection{Related Works}
Related works of joint waveform design focus on striking a balance between the tradeoff of communication and sensing. For example, \cite{liu2018toward} investigated the tradeoff between the multi-user interference minimization and the appropriate radar beampattern formulation. Besides, a recent work in~\cite{liu2021cram} considered the Cram\'{e}r-Rao bound (CRB) minimization with guaranteed signal-to-interference-plus-noise ratio (SINR) for each communication user. Furthermore, as widely-used performance metrics, the fundamental tradeoff between the CRB for target parameter estimation and the data rate for communication was also investigated in~\cite{hua2022mimo,ren2022fundamental} under various system settings, to unveil the potential of ISAC.

Despite the above approaches can achieve favorable performance tradeoffs between the estimation performance and spectral efficiency~\cite{hua2022mimo,ren2022fundamental,lushihang}, the energy efficiency (EE) optimization of the joint waveform has not been fully investigated. Currently, the energy consumption of the state-of-the-art fifth-generation (5G) wireless networks is extremely high, resulting in expensive operational costs~\cite{prasad2017energy,zappone2015energy2}. 
It is anticipated that the upcoming ISAC will pave the way for developing a perceptive wireless network requiring a much higher energy consumption than the current one, since the wireless signals are expected to achieve the dual purposes of environment sensing and information transmission simultaneously.    
This could hinder the long-term development of sustainable and environmentally friendly wireless communication technologies.
Hence, there is a pressing need to investigate the energy efficiency design of ISAC for establishing
a perceptive-efficient and spectrally-efficient cellular network.
Actually, energy-aware optimization has been a hot topic in the past decade for conventional cellular networks,
e.g.,~\cite{prasad2017energy,wang2015energy,zappone2015energy,zappone2015energy2}. 
Specifically, EE is defined as the ratio of the achieved data rate and the required power consumption, capturing the energy consumption per bit in communication, which has been widely studied for various communication networks~\cite{feng2012survey,tervo2015optimal}.
However, these approaches for maximizing the communication EE cannot be directly applied to ISAC, as they do not take into consideration of sensing functionalities.
Recently, the EE optimization for radar-communication spectrum sharing has been studied in ~\cite{grossi2021energy}, and the results cannot be applied to ISAC systems either due to the separated signal waveform design.
On the other hand, a few works have studied ISAC beamforming for  maximizing communication-centric EE. For instance, the work of \cite{he2022energy} investigated the communication EE maximization under the required radar beampattern constraint. Yet, it does not consider the sensing EE and the performance of target parameter estimation. Besides, the work of  \cite{huang2022integrated} focused on energy minimization under the sensing and communication constraints. In particular, the algorithm designed in~\cite{huang2022integrated} cannot handle the EE optimization due to the intrinsic challenges brought by fractional programming in the resource allocation design.  
More importantly, to the best of our knowledge, the sensing-centric EE that characterizes the EE of target sensing has been rarely studied in the literature.
In particular, to fulfill the increasing demand for sensing services, it is natural for the base station (BS) to transmit the waveforms with high power for improving the detection and estimation performance. However, this operation will inevitably bring unaffordable energy costs, which contradicts to the emerging requirements of carbon neutrality and environmental sustainability for future wireless networks~\cite{prasad2017energy}.
Therefore, there is an urgent need for the design an energy-efficient sensing performance metric for ISAC. 

\subsection{Contributions}

Against this background, this work considers the EE optimization for the waveform design of ISAC, where the communication-centric EE, sensing-centric EE, and their tradeoffs are investigated. 
Specifically, for the ISAC systems wherein communication serves as the primary objective, we study the ISAC waveform design for maximizing the communication-centric EE, i.e., the ratio of the achievable rate and the corresponding power consumption, while guaranteeing both the target estimation and communication performance in terms of the CRB and SINR, respectively.  
As for the sensing-centric ISAC systems, for the first time, we propose the performance metric to measure the sensing-centric EE for target parameter estimation. 
Then, we optimize the ISAC waveform to maximize the sensing-centric EE, considering the constraints of SINR, CRB, and the maximum transmission power budget. Then, we study the Pareto boundary of communication-centric EE and sensing-centric EE for characterizing their tradeoffs. The main contributions of this paper are summarized as follows.

\begin{itemize}
	\item We optimize the communication-centric EE considering the two scenarios having a point-like target estimation and an extended target estimation, respectively, under the constraints of CRB, SINR, and transmission power limitations. For the case of point-like target, the nonconvexity of the objective function and CRB constraint hinder the communication-centric EE optimization. For handling these challenges, we first adopt the quadratic-transform-Dinkelbach method to reformulate the nonconvex fractional objective function as a tractable formulation. Then, we adopt the semi-definite relaxation and linear matrix inequality to convert the nonconvex optimization problem into a sequence of convex optimization problems. Finally, we generalize the proposed algorithm to an extended target case.

    \item We propose a performance metric for capturing the notion of sensing-centric EE for the first time, which adopts the ratio of the reciprocal of the CRB to the transmit energy for measuring “\textit{information-per-Joule}’’. Then, based on the proposed metric, we consider the sensing-centric EE maximization for point-like/extended targets by optimizing the transmit beamforming. Although the considered problem is nonconvex, we adopt the Schur complement to reformulate the problem into a tractable formulation, facilitating the development of a successive convex approximation (SCA)-based algorithm to effectively acquire the solution to the design problem. 

    \item We adopt the Pareto optimization technique to characterize the tradeoff between the communication-centric EE and the sensing-centric EE. In particular, we formulate a constrained optimization problem that maximizes the communication-centric EE under the constraint of sensing-centric EE. To handle the nonconvexity of the considered optimization problem, we propose an SCA-based iterative algorithm for addressing the nonconvexity. Then, by varying the threshold of the sensing-centric EE, the approximate Pareto boundary can be obtained by solving a sequence of constrained problems. Simulation results present the Pareto boundary to demonstrate the tradeoff between the two EE metrics.  
 
\end{itemize}


The remainder of this paper is organized as follows. Section II introduces the system model, including the communication model and the sensing model. In Section III, we study the optimization of  the communication-centric EE under the sensing and communication constraints. The sensing-centric EE is studied in Section IV. Section V investigates the  tradeoff between the communication-centric and the sensing-centric EE. Simulation results are provided in Section VI. Finally, we conclude the paper in Section VII.

Notations: The normal plain text (i.e., $t$), bold lowercase letters (i.e., $\mathbf{w}$) and uppercase letters (i.e., $\mathbf{W}$) represent scalars, vectors, and matrices, respectively.  $\operatorname{tr}(\cdot)$, $\operatorname{rank}(\cdot)$, $(\cdot)^H$, and $(\cdot)^T$ denote the trace operator, the rank operator, the Hermitian transpose, and the transpose operator, respectively. $\mathbb{C}^{n \times n}$ stands for an $n \times n$ complex-valued matrix. $\| \cdot \|$ represents the $L_2$ norm of a matrix. The inequality $\mathbf{A} \succeq \mathbf{0}$ means that $\mathbf{A}$ is Hermitian positive semi-definite. $\operatorname{Re}(\cdot)$ denotes the real part of the argument. We adopt $\mathbb{E}(\cdot)$ for the stochastic expectation. $\dot{f}(x)$ denotes the first derivative of function $f(x)$. The notation $\triangleq$ is used for definitions.

\section{System Model}

\begin{figure}
	\centering
	\includegraphics[width=0.4\linewidth]{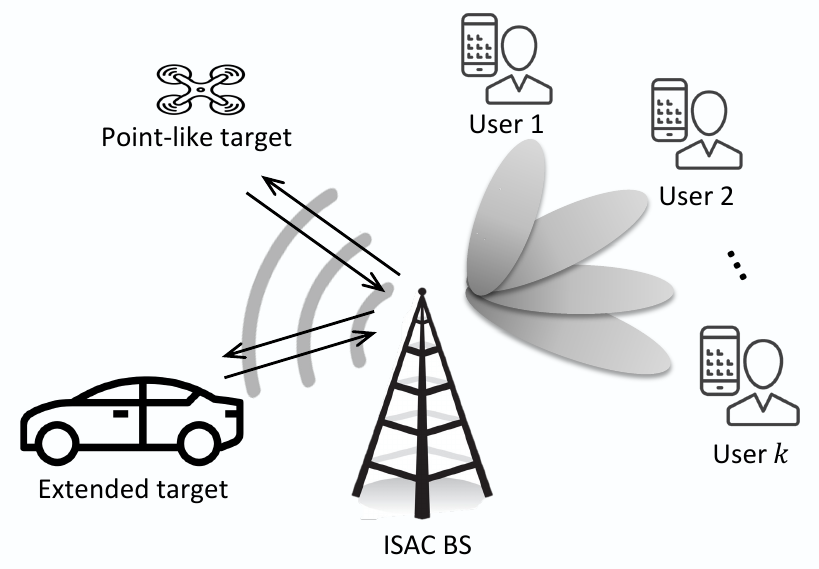}
	\caption{An illustration of an ISAC MIMO system where an ISAC BS simultaneously serves multiple communication users and senses a point-like target or an extended target.}
	\label{fig:fig1}
\end{figure}

As depicted in Fig. \ref{fig:fig1}, we consider an ISAC multiple-input multiple-output (MIMO) system, where the BS equipped with $M$ transmit antennas serves $K$ single-antenna UEs for communication with $K \leq M$. Let $k \in \mathcal{K}  \triangleq \{1,2, \cdots,K\}$ denote the communication user set.  As for radar estimation, the environmental information is simultaneously extracted from the reflected echoes with $N$ receiving antennas implemented at the BS. 
Without loss of generality, the number of transmit antennas is less than that of receive antennas, i.e., $M \le N$. As for target sensing, both the point-like target and the extended target cases are considered separately covering various practical scenarios. In particular, the former case denotes the unstructured point that is far away from the BS, such as unmanned aerial vehicles (UAVs). On the other hand, for the extended target, it acts as a reflecting surface with a large number of distributed scatterers, such as a vehicle or a pedestrian~\cite{liu2021cram}. The detailed model is given as follows.

\subsection{Communication Model}


We denote the beamforming vector and the channel from the BS to the $k$-th user as $\mathbf{w}_k\in\mathbb{C}^{M\times 1}$ and $\mathbf{h}_k\in\mathbb{C}^{M\times 1}$, respectively. Then, the data symbol intended for the $k$-th user at time slot $l$ is denoted as ${s}_k[l] $, with unit power $\mathbb{E} \left( |{s}_k[l]|^2\right)  =1$.  Left multiplying $\mathbf{s}[l] = [s_1[l], s_2[l], \cdots, s_k[l]]^T \in \mathbb{C}^{K \times 1}$ with the beamforming matrix $\mathbf{W} = [\mathbf{w}_1, \mathbf{w}_2, \cdots, \mathbf{w}_k] \in \mathbb{C}^{M \times K}$, the transmitted signal vector of the BS is given by $\mathbf{x}[l]= \mathbf{W} \mathbf{s}[l].$
Then, the transmitted ISAC waveform over $L$ time slots can be denoted as $\mathbf{X} =  [ {\bf x}[1], {\bf x}[2], \cdots,$ $ {\bf x}[L] ]   \in \mathbb{C}^{M \times L} $. Then, the received signal at the $k$-th user during the $l$-th time slot, $l \in \{1, 2, \cdots, L\}$, is given as follows
\begin{align}
	y_k[l] = {\bf h}_k^H \mathbf{w}_k s_k[l] + 
	\mathop{\sum}_{k \in \mathcal{K} \atop j \neq k} {\bf h}_k^H  \mathbf{w}_j s_j[l] + z_c[l], \label{eq:ykn}
\end{align}
where $z_c[l]$ is the additive white Gaussian noise (AWGN) with zero mean and variance $\sigma_c^2$. The received SINR at the $k$-th user can be calculated as 

\begin{align}
	\text{SINR}_k({\bf W}) = \frac{\left\vert {\bf h}_k^{H} {\bf w}_k \right\vert^2}{\sigma_c^2 + \mathop{\sum}_{k \in \mathcal{K} \atop j \neq k}\left\vert {\bf h}_k^H {\bf w}_j \right\vert^2}, \label{eq:sinr}
\end{align}
and the corresponding achievable rate is $R_k({\bf W}) = \log_2(1+\text{SINR}_k ({\bf W})). $

It is well known that communication-centric EE is defined as a ratio of the transmission sum rate $\sum_k R_k({\bf W})$ to the total power consumption $P$. Following~\cite{xu2011improving,arnold2010power}, the power consumption can be calculated as
\begin{align}
	P = \frac{1}{\epsilon}P_d + P_0,
\end{align}
where the power amplifier efficiency $\epsilon \in [0,1]$ and $P_0$ denotes the constant circuit power consumed by circuitries in RF chains, power supply, cooling system, etc. Besides, the total transmit power is given by $P_d = \sum_k \|{\bf w}_k\|_2^2$. Hence, the communication-centric EE, measuring the required ``bits-per-Joule"~\cite{tervo2015optimal,prasad2017energy}, can be calculated as
    \begin{align}
	\text{EE}_\mathrm{C} & = \frac{R_k(\mathbf{W})}{ P }  = \frac{\sum_k \log_2\left( 1+\left\vert {\bf h}_k^H {\bf w}_k \right\vert^2 \bigg/ \left( \sigma_c^2 + \mathop{\sum}_{k \in \mathcal{K} \atop j \neq k}\left\vert {\bf h}_k^H {\bf w}_j \right\vert^2\right)  \right) }{\frac{1}{\epsilon}\sum_k \|{\bf w}_k\|_2^2 + P_0}.  \label{EE}
\end{align}

\subsection{Sensing Model}

For radar sensing, the BS exploits the echo signals collected in $L$ time slots to estimate the target parameter. 
This work considers  the two cases with either a point-like target or an extended target, respectively.
For notational simplicity, we consider the same angle of departure (AOD) and angle of arrival (AOA) of the target, i.e., $\theta_t=\theta_r=\theta$~\cite{bekkerman2006target}. Then, 
for the point-like target that locates in the far field, the target response matrix can be denoted as
\begin{align}
  \mathbf{A} =  \alpha \mathbf{a}_r(\theta)\mathbf{a}^{H}_t(\theta), \label{target_response_point}
\end{align}
where $\mathbf{a}_x(\theta)$, $x\in\{t,r\}$, is the steering vector for the transmit signal at angle $\theta$. Following the existing works on ISAC, e.g., \cite{hua2022mimo,ren2022fundamental}, we assume that the BS employs a uniform linear antenna with a half-wavelength spacing between the adjacent antennas. Then, the transmit and receive steering vectors are given by
\begin{align}
    \mathbf{a}_t(\theta) =  \left[ 1, \cdots, e^{-j \pi cos\theta}, e^{-j \pi (M -1) cos\theta} \right]^T,\\
\mathbf{a}_t(\theta) =  \left[ 1, \cdots, e^{-j \pi cos\theta}, e^{-j \pi (N -1) cos\theta} \right]^T.
\end{align}

 For the extended target that locates in the near field, we follow \cite{liu2021cram} to model it as a reflecting surface with $N_s$ point-like scatters. Then, the target response matrix can be represented as 
\begin{align}
	{\mathbf{A}} =  \sum_{i=1}^{N_s} \alpha_i \mathbf{a}_r(\theta_i)\mathbf{a}_t^{H}(\theta_i), \label{target_response}
\end{align}
where $\alpha_i$ is the reflection coefficient of the $i$-th scatterer. 


Therefore, the received target echoes $\mathbf{Y}_\mathrm{R}$ from the point-like or the extended targets can both be denoted as
\begin{equation}
	\mathbf{Y}_\mathrm{R} = \mathbf{A} \mathbf{X} + \mathbf{Z}_s,  \label{eq:yn}   
\end{equation}
where $\mathbf{Z}_s$ is the zero-mean AWGN with variance $\sigma_s^2$ in each element.


Since CRB is a lower bound on the variance of an unbiased estimator of an unknown parameter that can guarantee the performance of sensing \cite{liu2021cram,ren2022fundamental}, we adopt the CRB as the sensing metric to design the energy-efficient ISAC in the following.

\section{Communication-Centric Energy-Efficient Design} \label{sec-comm}

\subsection{Point-Like Target Case} \label{sec-comm-point}
Since the CRB of $\alpha$ has a similar form as the one of $\theta$, for conciseness,
this work only considers the CRB of $\theta$ to for the design of the ISAC  beamforming. For the point-like target, the CRB of $\theta$ is given as follows~\cite{bekkerman2006target}
\begin{align}
	{\hbox{CRB}}(\theta)=\frac{\sigma_s^2}{ \left\vert \alpha \right\vert^2 \left(M\dot{\mathbf{a}}^{H}(\theta)\mathbf{R}_\mathbf{x}^{T}\dot{\mathbf{a}}(\theta)+ \mathbf{a}^{H}(\theta)\mathbf{R}_\mathbf{x}^{T} \mathbf{a}(\theta)\left\Vert\dot{\mathbf{a}}(\theta)\right\Vert^{2}-\frac{M\left\vert \mathbf{a}^{H}(\theta)\mathbf{R}_\mathbf{x}^{T}\dot{\mathbf{a}}(\theta)\right\vert^{2}}{ \mathbf{a}^{H}(\theta)\mathbf{R}_\mathbf{x}^{T} \mathbf{a} (\theta)}\right)},  \label{eq:CRB}
\end{align} 
where $\mathbf{R}_\mathbf{x}$ is the sample covariance matrix of $\mathbf{X}$. Since $\mathbb{E} \left( |{s}_k[l]|^2\right)  =1$, for a large $L$, we have the  asymptotic result
${\bf R}_\mathbf{x} = \frac{1}{L}{\bf X}{\bf X}^H \approx {\bf W} {\bf W}^H = \sum_{k=1}^K {\bf w}_k{\bf w}_k^H$ ~\cite{liu2021cram}.

The communication-centric energy efficient design is to maximize the EE$_\mathrm{C}$ defined in \eqref{EE}, under the constraints of multiple users’ required SINR and maximal CRB$(\theta)$, whose optimization problem can be formulated as follows
\begin{subequations}\label{original_problem}
	\begin{align}
		\max_{ \{\mathbf{w}_k\}_{k=1}^{K}}~~&\frac{\sum_{k=1}^K  \log_2 \left( 1+\left\vert {\bf h}_k^H {\bf w}_k \right\vert^2 \bigg/ \left( \sigma_c^2 + \mathop{\sum}_{k \in \mathcal{K} \atop j \neq k}\left\vert {\bf h}_k^H {\bf w}_j \right\vert^2\right)  \right) }{\frac{1}{\epsilon}\sum_k \|{\bf w}_k\|_2^2 + P_0} \label{eq:p1a}\\
		\mathrm{s.t.} ~~&\sum_{k=1}^K  \|{\bf w}_k \|_2^2 \leq P_\text{max}, \label{eq:p1b} \\
		&\text{CRB}(\theta) \leq \rho ,\label{eq:p1c} \\
		&	\frac{\left\vert {\bf h}_k^{H} {\bf w}_k \right\vert^2}{\sigma_c^2 + \mathop{\sum}_{k \in \mathcal{K} \atop j \neq k}\left\vert {\bf h}_k^H {\bf w}_j \right\vert^2} \geq \gamma_k, \forall k, \label{eq:p1d}
	\end{align} 
	\label{eq:p1}%
\end{subequations}
where $P_\text{max}$ denotes the power budget of the BS and \eqref{eq:p1b} is the transmit power constraint. 
Besides, $\rho$ and $\gamma_k$ are the required CRB threshold for sensing and the required SINR for the $k$-th communication user, respectively.
In general, it is challenging to solve problem \eqref{eq:p1} directly, due to the nonconvexity of the fractional objective function \eqref{eq:p1a} and  nonconvex constraints  \eqref{eq:p1c} and \eqref{eq:p1d}.
 
For addressing the nonconvex optimization problem, we first adopt the Dinkelbach's method \cite{NonlinearFractionalProgramming} to reformulate the problem \eqref{original_problem} as 
\begin{subequations}\label{reformulated_problem}
 \begin{align}
	\max_{ \{\mathbf{w}_k\}_{k=1}^{K}}~~& f_1(\mathbf{w}_k) - \lambda f_2(\mathbf{w}_k) \label{eq:dinkelbach_1}\\
\mathrm{s.t.}~~~&(\ref{eq:p1b}),\;(\ref{eq:p1c}),\;(\ref{eq:p1d}),\label{added_constraint}
\end{align}   
\end{subequations}
where $f_1(\mathbf{w}_k) \triangleq {\sum_{k=1}^K \log_2 \left( 1+\frac{\left\vert {\bf h}_k^H {\bf w}_k \right\vert^2}{\sigma_c^2 + \sum_{j=1,j \neq k}^K \left\vert {\bf h}_k^H {\bf w}_j \right\vert^2}  \right) }$,
$f_2(\mathbf{w}_k) \triangleq {\frac{1}{\epsilon} \sum_{k=1}^K \|{\bf w}_k\|_2^2 + P_0}$, and $\lambda \geq 0$ is the auxiliary variable to be iteratively updated by 
\begin{align}
\lambda = \frac{f_1(\mathbf{w}_k)}{f_2(\mathbf{w}_k)}.\label{lambda_update}
\end{align}
With \eqref{reformulated_problem} and \eqref{lambda_update}, an efficient solution to problem (\ref{original_problem}) can be obtained by updating $\mathbf{w}_k$ and $\lambda$ alternately. 

Nevertheless, problem (\ref{reformulated_problem}) is still difficult to handle due to the following issues: 1) the objective function \eqref{eq:dinkelbach_1} is still non concave over $\{ \mathbf{w}_k \}$ due to the fractional function $f_1(\mathbf{w}_k)$; 2) nonconvex constraints (\ref{eq:p1c}) and (\ref{eq:p1d}).
Since the function $\log_2(\cdot)$ is concave and non-decreasing, the nonconvexity of \eqref{eq:dinkelbach_1} can be addressed if the term inside $\log_2(\cdot)$ can be reformulated as an equivalent concave formulation.
Bearing this in mind, since $f_1(\mathbf{w}_k)$ belongs to the general multiple-ratio concave-convex fractional programming problem, we adopt the quadratic transform method~\cite[Theorem 1]{shen2018fractional} to reformulate $f_1(\mathbf{w}_k)$ as 
\begin{equation}
 f_1(\mathbf{w}_k) = \underset{t_k}{\mathrm{max}}\sum_{k=1}^K \log_2 \left( 1+ 2 t_k \operatorname{Re}(\mathbf{w}_k^H \mathbf{h}_k)
 - t_k^2 B_k(\mathbf{w}_k)  \right) , \label{eq:qt_1}
\end{equation}
where $B_k(\mathbf{w}_k) = \sigma_c^2 + \sum_{j=1,j \neq k}^K \left\vert {\bf h}_k^H {\bf w}_j \right\vert^2$ and $t_k$ is an introduced auxiliary variable that is iteratively updated by
\begin{equation}
t_k = \left\vert {\bf h}_k^{H} {\bf w}_k \right\vert  \left( \sigma_c^2 +\sum_{j=1,j \neq k}^K\left\vert {\bf h}_k^H {\bf w}_j \right\vert^2\right)^{-1}.
\end{equation}
Based on the above reformulations, problem (\ref{original_problem}) can be recast as 
\begin{align}  
	\max_{ \{\mathbf{w}_k, t_k\}_{k=1}^{K}, \lambda}~~& \sum_{k=1}^K \log_2\left( 1+ 2 t_k \operatorname{Re}(\mathbf{w}_k^H \mathbf{h}_k) - t_k^2 B_k(\mathbf{w}_k)  \right)  -  \lambda \left(  {\frac{1}{\epsilon} \sum_{k=1}^K \|{\bf w}_k\|_2^2 + P_0} \right) ~~~ 
 \mathrm{s.t.}& (\ref{added_constraint}),
\end{align}
where $\{\mathbf{w}_k, t_k\}_{k=1}^{K}$ and $\lambda$ can be updated alternatively. 

In the following, we focus on handling the nonconvex constraints \eqref{eq:p1c} and \eqref{eq:p1d}. Specifically,  constraint (\ref{eq:p1c}) can be reformulated as
\begin{align}
	& M\dot{\bf a}^{H}(\theta){\bf R}_{\bf x}^{T}\dot{\bf a}(\theta)+{\bf a}^{H}(\theta){\bf R}_{\bf x}^{T}{\bf a}(\theta)\left\Vert\dot{\bf a}(\theta)\right\Vert^{2}  -  \frac{M\left\vert{\bf a}^{H}(\theta){\bf R}_{\bf x}^{T}\dot{\bf a}(\theta)\right\vert^{2}}{{\bf a}^{H}(\theta){\bf R}_{\bf x}^{T}{\bf a}(\theta)} - \frac{\sigma_s^2}{2L\rho\left\vert \alpha \right\vert^2 }\geq 0.\label{eq:p1c_1}
\end{align}
Then, for notational conciseness, denoting $\mathcal{F}({\bf R_X}) \triangleq M\dot{\bf a}^{H}(\theta){\bf R}_{\bf x}^{T}\dot{\bf a}(\theta)+{\bf a}^{H}(\theta){\bf R}_{\bf x}^{T}{\bf a}(\theta)\left\Vert\dot{\bf a}(\theta)\right\Vert^{2} $, \eqref{eq:p1c_1} can be reformulated as the following linear matrix inequality by leveraging the Schur complement ~\cite{zhang2006schur}.
\begin{align}
	\begin{bmatrix}
		\mathcal{F}({\bf R_x}) - \frac{\sigma_s^2}{2L\rho\left\vert \alpha \right\vert^2}      & \sqrt{M}{\bf a}^{H}(\theta){\bf R}_{\bf x}^{T}\dot{\bf a}(\theta)      \\
		\sqrt{M}\dot{\bf a}^{H}(\theta){\bf R}_{\bf x}^{T}{\bf a}(\theta)     & {\bf a}^{H}(\theta){\bf R}_{\bf x}^{T}{\bf a}(\theta) 
	\end{bmatrix} \succeq \mathbf{0} .  \label{eq:p1c_2}
\end{align}
Next, for handling the nonconvex constraint \eqref{eq:p1d}, we introduce an auxiliary optimization variable matrix $\mathbf{W}_k$ and reformulate  constraint \eqref{eq:p1d} into 
\begin{subequations}
	\begin{align}
 \text{tr}(\mathbf{Q}_k \mathbf{W}_k) - \gamma_k \mathop{\sum}_{k \in \mathcal{K} \atop j \neq k} \text{tr}(\mathbf{Q}_k \mathbf{W}_j) &\geq \gamma_k \sigma_c^2,\label{eq:Www1} \\
 {\bf W}_k & =\textbf{w}_k{\bf w}_k^H, \label{eq:Www}
	\end{align}
\end{subequations}
where $\mathbf{Q}_k = {\bf h}_k{\bf h}_k^H$. Then, problem (\ref{original_problem}) can be equivalently reformulated as

\begin{subequations}
\begin{align}
\max_{\{\mathbf{w}_k,\mathbf{W}_k, t_k\}_{k=1}^{K}}~~&{\sum_{k=1}^K \log_2 \left( 1+ 2 t_k \cdot \operatorname{Re}(\mathbf{w}_k^H \mathbf{h}_k) - t_k^2 B_k(\mathbf{W}_k)  \right)  } - \lambda \left( {\frac{1}{\epsilon} \sum_{k=1}^K \text{tr}({\bf W}_k)+ P_0}\right)  \\
\mathrm{s.t.} ~~~& \begin{smallmatrix}\begin{bmatrix}
 		\mathcal{F}(\sum_{k=1}^K\mathbf{W}_k) - \frac{\sigma_s^2}{2L\rho\left\vert \alpha \right\vert^2}      & \sqrt{M}{\bf a}^{H}(\theta)\sum\limits_{k=1}^K{\bf W}_k^T\dot{\bf a}(\theta)      \\
 		 \sqrt{M}\dot{\bf a}^{H}(\theta)\sum\limits_{k=1}^K{\bf W}_k^T{\bf a}(\theta)     & {\bf a}^{H}(\theta)\sum\limits_{k=1}^K{\bf W}_k^T{\bf a}(\theta)    
 	\end{bmatrix}\end{smallmatrix} \succeq \mathbf{0} ,\label{matrix_inequality_constraint_added1}\\
&(\ref{eq:p1b}),\;(\ref{eq:Www1}),\;(\ref{eq:Www}),
\end{align}    
\end{subequations}
where $B_k(\mathbf{W}_k) \triangleq \mathop{\sum}_{k \in \mathcal{K} \atop j \neq k} \text{tr}(\mathbf{Q}_k \mathbf{W}_j) + \sigma_c^2$. However, constraint (\ref{eq:Www}) is a nonconvex equality constraint which is difficult to handle. Therefore, we introduce the following lemma to transform constraint (\ref{eq:Www}) into equivalent inequality constraints.

\begin{Lm} \label{lm:Www}
${\bf W}_k =\textbf{w}_k{\bf w}_k^H$ can be equivalently reformulated as 
\begin{subequations}
	\begin{align}
		& \begin{bmatrix}
			\mathbf{W}_k      & \mathbf{w}_k       \\
			\mathbf{w}_k^H     & 1    
		\end{bmatrix} \succeq \mathbf{0} , \mathbf{W}_k \succeq \mathbf{0}, \forall k, \label{eq:rank1a} \\
		& \operatorname{tr}(\mathbf{W}_k) - \mathbf{w}^H_k \mathbf{w}_k \leq 0, \forall k. \label{eq:rank1b}
	\end{align}
\end{subequations}
\end{Lm}
\begin{IEEEproof}
 The proof is given in Appendix A.
\end{IEEEproof}

Although the equality constraint in (\ref{eq:Www}) has been reformulated as the equivalent inequality constraints, constraint (\ref{eq:rank1b}) is still nonconvex.
For handling this, we  adopt the SCA technique that establishes an inner convex approximation of constraint (\ref{eq:rank1b}) given as 
\begin{align}
    \operatorname{tr}(\mathbf{W}_k) + \left({\mathbf{w}}_k^{(i-1)}\right)^H {\mathbf{w}}_k^{(i-1)} - 2\operatorname{Re}\left(\left({\mathbf{w}}_k^{(i-1)}\right)^H \mathbf{w}_k \right) \leq 0, \forall k, \label{inequality_constraint_added}
\end{align}
where ${\mathbf{w}}^{(i-1)}_k$ is the solution obtained at the $i$-th iteration of the SCA.

Therefore, at the $i$-th iteration, the convex approximation of problem \eqref{eq:p1} can be reformulated as
 \begin{subequations}\label{approximateproblem1}
\begin{align}  
	\max_{ \mathcal{W}, t_k, \lambda}~~&{\sum_{k=1}^K  \log_2 \left( 1+ 2 t_k \operatorname{Re}(\mathbf{w}_k^H \mathbf{h}_k) - t_k^2 B_k(\mathbf{W}_k)  \right)  } -  \lambda \left( {\frac{1}{\epsilon} \sum_{k=1}^K \text{tr}({\bf W}_k)+ P_0}\right)   \label{eq:p1_convex}\\
	\mathrm{s.t.}  ~~
	&(\ref{eq:p1b}), (\ref{eq:Www1}),(\ref{matrix_inequality_constraint_added1}),(\ref{eq:rank1a}),(\ref{inequality_constraint_added}).   \label{eq:p1_convex_constraints}
\end{align}
 \end{subequations}

 Algorithm~\ref{alg1} summarizes the iterative algorithm for handling  problem \eqref{original_problem}, where $\hat{f}_1(\mathbf{w}_k, \mathbf{W}_k) = \sum_{k=1}^K  \log_2\left( 1+ 2 t_k \operatorname{Re}(\mathbf{w}_k^H \mathbf{h}_k) - t_k^2 B_k(\mathbf{W}_k)  \right) $ and $\hat{f}_2(\mathbf{W}_k) ={\frac{1}{\epsilon} \sum_{k=1}^K \text{tr}({\bf W}_k)+ P_0} $. Although we cannot guarantee that the optimal solution of problem \eqref{eq:p1} can be obtained,  the proposed Algorithm~\ref{alg1} follows the inexact Dinkelbach-type algorithm adopted in \cite{InexactFF}, whose convergence can be guaranteed by the following lemma.
 \begin{Lm}
 Let $\left\{\mathbf{w}_k^{i},\mathbf{W}_k^i\right\}$ be the solution sequence generated by solving problem \eqref{approximateproblem1}. The sequence $\{\lambda^{(i)}\}$ generated by Algorithm 1 is non-decreasing and convergent.
 \end{Lm}
 \begin{IEEEproof}
  Since
$\hat{f}_1(\mathbf{w}^{(i)},\mathbf{W}^{(i)})-\lambda^{(i)}{\hat{f}_2(\mathbf{W}^{(i)})}
=\left(\lambda^{(i+1)}-\lambda^{(i)}\right){\hat{f}_2(\mathbf{W}^{(i)})}$,
 we have $\lambda^{(i+1)}\geq\lambda^{(i)}$ if $\hat{f}_1(\mathbf{w}^{(i)},\mathbf{W}^{(i)})-\lambda^{(i)}{\hat{f}_2(\mathbf{W}^{(i)})}\geq 0$. 
Obviously, $\hat{f}_1(\mathbf{w}^{(i-1)},\mathbf{W}^{(i-1)})-\lambda^{(i)}{\hat{f}_2(\mathbf{W}^{(i-1)})}=0$. At the $i$-th iteration, we approximate problem \eqref{original_problem} as
problem \eqref{approximateproblem1}  around $\mathbf{w}_k^{(i-1)}$.  Since $\mathbf{w}_k^{(i-1)}$ is definitely a feasible solution of problem \eqref{approximateproblem1}, we have
$\hat{f}_1(\mathbf{w}^{(i)},\mathbf{W}^{(i)})-\lambda^{(i)}{\hat{f}_2(\mathbf{W}^{(i)})}\geq \hat{f}_1(\mathbf{w}^{(i-1)},\mathbf{W}^{(i-1)})-\lambda^{(i)}{\hat{f}_2(\mathbf{W}^{(i-1)})}= 0$.
Therefore, we can conclude that the sequence $\{\lambda^{(i)}\}$ is non-decreasing and Algorithm 1 converges due to the finite power budget. 
 \end{IEEEproof}
\begin{algorithm} 
	\caption{\it: Proposed Iterative Algorithm for Handling \eqref{original_problem}  } 
	\label{alg1} 
	\begin{algorithmic}
		\STATE Set $i = 0$, $\delta > 0$,$ \{ \mathbf{w}_k^{(0)}, \mathbf{W}_k^{(0)} \} \in \mathcal{S};$
		\STATE Initialize $t_k^{(0)}, \lambda^{(0)}$ satisfying $\hat{f}_1(\mathbf{w}^{(0)},\mathbf{W}^{(0)})-\lambda^{(0)}{\hat{f}_2(\mathbf{W}^{(0)})} \ge 0;$ 
		\REPEAT 
		\STATE $i \leftarrow i + 1$ ;
        \STATE $\tilde{\mathbf{w}}_k^{(i)} \leftarrow \mathbf{w}_k^{(i-1)}$;
		\STATE $ t^{(i)}_k \leftarrow \frac{\operatorname{Re}({\mathbf{w}_k^{(i-1)}}^H \mathbf{h}_k)}{B_k(\mathbf{W}_k^{(i-1)})}$;
		\STATE $\lambda^{(i)} \leftarrow \frac{\hat{f}_1(\mathbf{w}_k^{(i-1)}, \mathbf{W}_k^{(i-1)})}{\hat{f}_2(\mathbf{W}_k^{(i-1)})}$ ;
		\STATE Solve problem \eqref{eq:p1_convex} to obtain the optimal $\mathbf{w}_k^{(i)}, \mathbf{W}_k^{(i)}$ ;
		\UNTIL $\hat{f}_1(\mathbf{w}^{(i)},\mathbf{W}^{(i)})-\lambda^{(i)}{\hat{f}_2(\mathbf{W}^{(i)})}$ is less than $\delta$.   
	\end{algorithmic}
\end{algorithm}

\textit{Complexity Analysis}:
The computational complexity of Algorithm~\ref{alg1} is dominated by solving problem \eqref{eq:p1_convex}. Problem \eqref{eq:p1_convex} involves linear matrix inequality (LMI) constraints that dominate the computation complexity. We notice that the problem contains one LMI constraint of size $2M$, $K$ LMI constraints of size $M+1$, and $K$ LMI constraints of size $M$. 
Given the required accuracy $\epsilon_0 > 0$, the  $\epsilon_0$-optimal solution can be achieved after a sequence of iterations.  Then, the computational complexity can be given as $\mathcal{O}( \sqrt{(2M +1)(K+1)} M^6 K^3$ $I_\mathrm{iter} \ln(1/\epsilon_0) )$ by reserving the highest order term, where $I_\mathrm{iter}$ denotes the number of iterations~\cite{ben2001lectures}.

\begin{propRemark}
Due to the stringent requirement introduced by \eqref{inequality_constraint_added}, it is generally non-trival to directly obtain a feasible solution as an initial point. Alternatively, we can adopt the penalty SCA~\cite{wang2021intelligent} and introduce auxilary variables $\bar{\rho}_k$ to transform problem \eqref{approximateproblem1} into

 \begin{subequations} \label{penaltySCA}
	\begin{align}  
		\max_{ \mathcal{W}, t_k, \lambda}~~&{\sum_{k=1}^K  \log_2 \left( 1+ 2 t_k \operatorname{Re}(\mathbf{w}_k^H \mathbf{h}_k) - t_k^2 B_k(\mathbf{W}_k)  \right)  } -  \lambda \left( {\frac{1}{\epsilon} \sum_{k=1}^K \text{tr}({\bf W}_k)+ P_0}\right) - \bar{p} \sum_{k=1}^K \bar{\rho}_k  \\
		\mathrm{s.t.}  ~~
		& {\mathbf{w}}_k^{(i-1)} - 2\operatorname{Re}\left(\left({\mathbf{w}}_k^{(i-1)}\right)^H \mathbf{w}_k \right) \leq \bar{\rho}_k, \forall k, \label{eq:addpenalty} \\
	&(\ref{eq:p1b}), (\ref{eq:Www1}), (\ref{matrix_inequality_constraint_added1}), (\ref{eq:rank1a}),   
	\end{align}
\end{subequations}
where $\bar{p}$ and $ \sum_{k=1}^K \bar{\rho}_k$ denote the weight coefficient and the penalty term, respectively.  To obtain the initial point of \eqref{approximateproblem1}, we can solve problem \eqref{penaltySCA} as an initial warm-up phase by gradually raising $\bar{p}$ to induce a reduction in the penalty term to a smaller value. When the penalty term decreases to zero, problem \eqref{penaltySCA} reduces to problem \eqref{approximateproblem1}, whose solution serves as the feasible initial point of \eqref{approximateproblem1}.
\end{propRemark}

\subsection{Extended Target Case} \label{sec:com-extend}
For estimating the extended target, we follow~\cite{liu2021cram} to consider the CRB of the target response matrix	$\mathbf{A}$ instead of the angle. Since $K \leq M$, transmitting $K$ signal streams is not always sufficient for recovering the rank-$M$ matrix. To address this issue, the BS generates additional signals that are dedicated for target probing. As such, the augmented data matrix at the $l$-th time slot is $\tilde{\mathbf{x}}[l]\triangleq\left[\mathbf{W}, \tilde{\mathbf{W}}\right]\left[\mathbf{s}[l];\tilde{\mathbf{s}}[l]\right]$, where $\tilde{\mathbf{s}}[l] \in \mathbb{C}^{(N_t-K) \times 1 }$ is the dedicated probing signal and $\mathbb{E} \left( \mathbf{s}[l] \tilde{\mathbf{s}}^H[l] \right)  = \mathbf{0}$.
Note that in the augmented signal, the beamforming $\mathbf{W} = \left[\mathbf{w}_1, \mathbf{w}_2, \cdots, \mathbf{w}_{K} \right] \in \mathbb{C}^{M \times K}$ broadcasts the information data to the $K$ users and the beamforming $\tilde{\mathbf{W}} = \left[\mathbf{w}_{K+1}, \cdots, \mathbf{w}_{K+M} \right] \in \mathbb{C}^{M \times M} $ is employed to generate probing signals for enabling the estimation of the target response matrix. However, the introduced probing signals $\tilde{\mathbf{s}}[l]$ inevitably generate undesired interference to the served multiple users that introduces non-trivial tradeoff between sensing and communication. In particular, the  SINR received at the $k$-th user is given by
\begin{align} 
\tilde{\text{SINR}}_k = \frac{\left| {\mathbf{h}_k^H \mathbf{w}_k} \right|^2}{\sum \nolimits _{i = 1,i \ne k}^{K} {{{\left| {{\mathbf {h}}_k^H{{\mathbf {w}}_i}} \right|}^2} + \left\Vert {\mathbf {h}}_k^H{ \tilde{\mathbf{W}}} \right\Vert_2^2 + \sigma _C^2} }, \label{eq:sinr-extended}
\end{align}
where $ \left\Vert {\mathbf {h}}_k^H{ \tilde{\mathbf {W}}} \right\Vert^2_2$ is the additional interference due to the probing signals. 
In such a case, the CRB for the extended target estimation can be derived as
\begin{align}
	{\hbox{CRB}}_{\text{extended}}= \frac{\sigma_s^2 M}{N} \operatorname{tr} \left({{\mathbf {R_x}}^{ - 1}} \right), \label{eq:CRB_extended} 
\end{align}
where ${{\mathbf {R}}_\mathbf{X}} = \mathbf{W} \mathbf{W}^H +  \tilde{\mathbf{W}} \tilde{\mathbf{W}}^H  .$

%

Based on the discussions above, the problem of communication-centric EE optimization for estimating an extended target can be formulated as
\begin{subequations}
	\begin{align}
		\max_{ \{\mathbf{w}_k\}_{k=1}^{K+M}}~~~& \frac{\sum_{k=1}^K \log_2(1+{\tilde{\text{SINR}}} _k)}{\frac{1}{\epsilon} \sum_{k=1}^{K+M} \|{\bf w}_k \|_2^2  + P_0} \label{eq:p2a}\\
		\mathrm{s.t.}  ~~~&\sum_{k=1}^{K+M} \|{\bf w}_k \|_2^2 \leq P_{\text{max}}, \label{eq:p2b} \\
		&	{\hbox{CRB}}_{\text{extended}}= \frac{\sigma_s^2 M}{L} \operatorname{tr} \left({{\mathbf {R_x}}^{ - 1}} \right) \leq \tau ,\label{eq:p2c} \\
		& \tilde{\text{SINR}_k} \geq \gamma_k. \label{eq:p2d}
	\end{align}
	\label{eq:p2}
\end{subequations}
\vspace{-0.8cm}

Obviously,  although constraints \eqref{eq:p2b} and \eqref{eq:p2c} are both convex, the fractional objective function \eqref{eq:p2a} 
is still nonconvex. 
Following  Section~\ref{sec-comm-point}, we first adopt Dinkelbach’s transformation to handle the nonconvex fractional programming and reformulate the problem as follows
\begin{subequations} \label{eq:p2dinkelbach}
\begin{align} 
    	\max_{ \{\mathbf{w}_k\}_{k=1}^{K+M}}~~~~& {\sum_{k=1}^K \log_2 (1+{\tilde{\text{SINR}}} _k)} - \lambda \left( \frac{1}{\epsilon}\sum_{k=1}^{K+M} \|{\bf w}_k \|_2^2 + P_0\right)\\
		s.t. ~~~~&\eqref{eq:p2b}, \eqref{eq:p2c}, \eqref{eq:p2d}.
\end{align}
\end{subequations}
Then, by exploiting the equality $-\log a = \underset{b}{\max} (\log b - ab)$~\cite{shi2015secure}, problem \eqref{eq:p2dinkelbach} can be reformulated as
\begin{subequations}\label{addedproblem1}
\begin{align}
	\max_{ \{\mathbf{w}_k\}_{k=1}^{K+M}, \{b_k\}_{k=1}^{K}, \lambda} ~~~&\sum_{k=1}^K \log_2 \left( \left| {\mathbf{h}_k^H \mathbf{w}_k} \right|^2 + \sum_{i = 1,i \ne k}^{K} {{{\left| {{\mathbf {h}}_k^H{{\mathbf {w}}_i}} \right|}^2} + \left\Vert {\mathbf {h}}_k^H{ \tilde{\mathbf {W}}} \right\Vert_2^2 + \sigma _C^2} \right) \\
	&+ \sum_{k=1}^K\left(  \log_2 b_k - b_k ( \sum_{i = 1,i \ne k}^{K} \left| \mathbf{h}_k^H{{\mathbf {w}}_i} \right|^2 + \left\Vert \mathbf{h}_k^H \tilde{\mathbf{W}} \right\Vert_2^2 + \sigma _C^2 ) \right) \\
 &- \lambda \left( \frac{1}{\epsilon}\sum_{k=1}^{K+M} \|{\bf w}_k \|_2^2 + P_0\right) \notag 
 \\
	\mathrm{s.t.}  ~~~~&\eqref{eq:p2b}, \eqref{eq:p2c}, \eqref{eq:p2d}.
\end{align}
\end{subequations}
For obtaining a tractable formulation, by introducing auxiliary variables  $\mathbf{W}_k \triangleq \mathbf{w}_k \mathbf{w}_k^H, k \in [1, 2, \cdots, K]$ and $\mathbf{R}_{\tilde{\mathbf{W}}} = \tilde{\mathbf{W}}  \tilde{\mathbf{W}}^H$,  problem (\ref{addedproblem1}) can be reformulated  as

\begin{subequations} \label{eq:p2all}
	\begin{align} 
		\max_{ \{ \mathbf{W}_k, b_k\}_{k=1}^{K}, \mathbf{R_W}_2, \lambda}~~&{\sum_{k=1}^K \log_2 \left( \mathbf{h}_k^H \left( \mathbf{W}_k +\sum_{i = 1,i \ne k}^{K} \mathbf{W}_i + \mathbf{R}_{\tilde{\mathbf{W}}}  + \sigma _C^2 \right)  \mathbf{h}_k    \right)  }  \notag \\
		&+ \sum_{k=1}^K\left(  \log_2 b_k - b_k \left( \sum_{i = 1,i \ne k}^{K}  \mathbf{h}_k^H{{\mathbf {W}}_i}\mathbf{h}_k  + \mathbf{h}_k^H \mathbf{R}_{\tilde{\mathbf{W}}} \mathbf{h}_k + \sigma _C^2 \right) \right)  \notag  \\
  &-  \lambda \left( {\frac{1}{\epsilon}  \text{tr}\left( \sum_{k=1}^{K} \mathbf{W}_k  + \mathbf{R}_{\tilde{\mathbf{W}}} \right) + P_0}\right) 
  ,  \label{objective_function_add}\\
		\mathrm{s.t.}  ~& \text{tr}\left( \sum_{k=1}^{K} \mathbf{W}_k  + \mathbf{R}_{\tilde{\mathbf{W}}} \right)   \leq P_\text{max},  \label{eq:p2convexb} \\
		&\frac{\sigma_s^2 M}{N} \text{tr}\left( \left( \sum_{k=1}^{K} \mathbf{W}_k  + \mathbf{R}_{\tilde{\mathbf{W}}} \right) ^{-1} \right)  \leq \tau , \label{eq:p2convexc} \\
		& \mathbf{h}_k \mathbf{W}_k \mathbf{h}^H_k - \gamma_k \left(   \sum_{i = 1,i \ne k}^{K} \mathbf{h}_k^H \mathbf{W}_i \mathbf{h}_k +  \mathbf{h}_k^H  \mathbf{R}_{\tilde{\mathbf{W}}} \mathbf{h}_k \right)   \geq \gamma_k \sigma_c^2,  \label{eq:p2convexd} \\
		& \mathbf{W}_k \succeq \mathbf{0},  \forall k,  \mathbf{R}_{\tilde{\mathbf{W}}} \succeq \mathbf{0}, \label{eq:p2convexe} \\
		& \operatorname{rank}(\mathbf{W}_k) = 1, \forall k. \label{eq:p2convexf} 
	\end{align} \label{eq:p2convex}%
\end{subequations}%
After inspecting problem (\ref{eq:p2all}), we can find that all constraints are convex, except for constraint (\ref{eq:p2convexf}). Besides, the objective function in (\ref{objective_function_add}) includes three sets of optimization variables: $\left\{\lambda\right\}$, $\left\{b_k\right\}$, and $\left\{\{ \mathbf{W}_k\}_{k=1}^{K}, \mathbf{R}_{\tilde{\mathbf{W}}}\right\}$. Moreover, when fixing the other two sets, the objective function is convex with respect to the remaining one. Therefore, we first adopt the rank relaxation to remove constraint (\ref{eq:p2convexf}) and then  employ an alternating optimization (AO) algorithm to optimize three sets of optimization variables alternately. 
The detailed algorithm is summarized in Algorithm 2, where we denote 
\begin{subequations} 
\begin{align}
     \tilde{f}_1(\mathbf{W}_k, \mathbf{R}_{\tilde{\mathbf{W}}} ) =&{\sum_{k=1}^K \log_2 \left( \mathbf{h}_k^H \left( \mathbf{W}_k +\sum_{i = 1,i \ne k}^{K} \mathbf{W}_i + \mathbf{R}_{\tilde{\mathbf{W}}} + \sigma _C^2 \right)  \mathbf{h}_k    \right)  } \notag \\
    & + \sum_{k=1}^K\left(  \log_2 b_k - b_k \left( \sum_{i = 1,i \ne k}^{K}  \mathbf{h}_k^H{{\mathbf {W}}_i}\mathbf{h}_k  + \mathbf{h}_k^H \mathbf{R}_{\tilde{\mathbf{W}}} \mathbf{h}_k + \sigma _C^2 \right) \right)  \\
    \tilde{f}_2(\mathbf{W}_k, \mathbf{R}_{\tilde{\mathbf{W}}} ) = & \frac{1}{\epsilon}  \text{tr}\left( \sum_{k=1}^{K} \mathbf{W}_k  + \mathbf{R}_{\tilde{\mathbf{W}}}  \right) + P_0.
\end{align}
\end{subequations}

\begin{algorithm} 
	\caption{{\it: Proposed Iterative Algorithm for Handling \eqref{eq:p2}}}
	\label{alg1_ex} 
	\begin{algorithmic}
         \STATE Set $i = 0$, $\delta > 0$,$ \{ \mathbf{W}_k^{(0)},  \mathbf{R}_{\tilde{\mathbf{W}}}^{(0)}\} \in \mathcal{S}$, and initialize $\left(b_k^{(0)},\lambda^{(0)}\right)$ satisfying $\tilde{f}_1(\mathbf{W}^{(0)}_k, \mathbf{R}_{\tilde{\mathbf{W}}}^{(0)} ) -\lambda^{(0)} \tilde{f}_2(\mathbf{W}^{(0)}_k, \mathbf{R}_{\tilde{\mathbf{W}}}^{(0)} ) \ge 0$;
		\REPEAT 
		\STATE $i \leftarrow i + 1$ ;
        \STATE $\tilde{\mathbf{W}}_k^{(i)}, \mathbf{R}_{\tilde{\mathbf{W}}}^{(i)} \leftarrow \mathbf{W}_k^{(i-1)}, \mathbf{R}_{\tilde{\mathbf{W}}}^{(i-1)}$;
		\STATE $ b^{(i)}_k \leftarrow b^{(i-1)}_k$;
		\STATE $\lambda^{(i)} \leftarrow \frac{\tilde{f}_1(\mathbf{W}^{(i-1)}_k, \mathbf{R}_{\tilde{\mathbf{W}}}^{(i-1)} ) }{\tilde{f}_2(\mathbf{W}^{(i-1)}_k, \mathbf{R}_{\tilde{\mathbf{W}}}^{(i-1)} )}$ ;
		\STATE Solve the rank relaxation of problem \eqref{eq:p2all} to obtain the optimal $\mathbf{W}_k^{(i)},  \mathbf{R}_{\tilde{\mathbf{W}}}^{(i)}, b^{(i)}_k$ ;
		\UNTIL $\tilde{f}_1(\mathbf{W}^{(i)}_k, \mathbf{R}_{\tilde{\mathbf{W}}}^{(i)} )-\lambda^{(i)}\tilde{f}_2(\mathbf{W}^{(i)}_k, \mathbf{R}_{\tilde{\mathbf{W}}}^{(i)} )$ is less than $\delta$.   
	\end{algorithmic}
\end{algorithm}

In the following theorem, we will show that the rank-1 solution of problem (\ref{eq:p2all}) can be recovered from the solution generated by Algorithm 2. 
\begin{Th}
Given the optimal solution obtained by Algorithm~\ref{alg1_ex} as $\{\mathbf{W}_k^\ast, \mathbf{R}^\ast_{\tilde{\mathbf{W}}} \}$. When $K = 1$, 
\begin{align}
\hat{\mathbf{W}}^\ast = \frac{\mathbf{W}^\ast \mathbf{h}_k  \mathbf{h}_k^H \mathbf{W}^\ast}{ \mathbf{h}_k^H \mathbf{W}^\ast \mathbf{h}_k},  ~~ \hat{\mathbf{R}}^\ast_{\tilde{\mathbf{W}}}=  \mathbf{R}^\ast_{\tilde{\mathbf{W}}}
\end{align}
is  the optimal rank-1 solution that achieves identical performance as $\{\mathbf{W}_k^\ast, \mathbf{R}^\ast_{\tilde{\mathbf{W}}} \}$.
When $K > 1$, one can always construct the optimal solution that satisfies the rank-1 constraint acquiring the same performance.
\end{Th}
\begin{IEEEproof}
  The proof is given in  Appendix B.
\end{IEEEproof}

\textit{Complexity Analysis}:
We provide the computational complexity of Algorithm~\ref{alg1_ex} as follows. Similarly, the problem \eqref{eq:p2all} is a semidefinite program that can be solved by the standard interior-point algorithm. We note that the problem involves $K+1$ LMI constraints of size $M$. We  consider the highest order term and express the computational complexity as $\mathcal{O}( \sqrt{MK+M+K+1} M^6 K^3 I_\mathrm{iter} \log(1/\epsilon_0) )$ for an $\epsilon_0$-optimal solution, where $I_\mathrm{iter}$ represents the number of iterations~\cite{ben2001lectures}.

\section{Sensing-Centric Energy-Efficient Design}
\subsection{Performance Metric for Sensing-Centric EE}


It is well known that CRB is the inverse of Fisher information for the unbiased estimator~\cite{li2012interpretation, pakrooh2015analysis}. In fact, Fisher information is the statistical expected value of the observed information about an observable random variable. Considering these, we adopt the reciprocal ratio of the CRB to the transmit power, further normalized by the total time slot length. In this context, we arrive at a novel sensing-centric EE metric that measures the average sensing information per Joule, defined as 
\begin{align}
\mathrm{EE}_\mathrm{s} \triangleq \frac{\text{CRB}^{-1}}{L \left(  \frac{1}{\epsilon}\sum_{k=1}^K  \|{\bf w}_k\|_2^2 + P_0 \right) }. \label{eq:definition}
\end{align} 
In this manner, both the sensing-centric EE and communication-centric EE  measure the ``information'' per Joule, but the ``information'' has different meanings.

Based on the above metric, we study the waveform design to maximize the sensing-centric EE considering the point-like target and the extended target in Sections \ref{sec:sensing-p} and \ref{sec:sensing-e}, respectively.

\subsection{Point-Like Target Case} \label{sec:sensing-p}
Considering the point-like target, with the CRB of estimating $\theta$ given in \eqref{eq:CRB}, the sensing-centric EE optimization problem can be formulated as
\begin{subequations} \label{eq:p3}
	\begin{align}
			\max_{ \{\mathbf{w}_k\}_{k=1}^{K}}~~~&\frac{\text{CRB}^{-1}(\theta)}{ L \left(  \frac{1}{\epsilon}\sum_{k=1}^K  \|{\bf w}_k\|_2^2 + P_0 \right) } \label{eq:p3a} \\ 
			\mathrm{s.t.} ~~~&\sum_{k=1}^K \|{\bf w}_k \|_2^2 \leq P_\text{max}, \label{eq:p3b}  \\ 
			&\text{CRB}(\theta) \leq \rho , \label{eq:p3c} \\ 
			&\frac{\left\vert {\bf h}_k^{H} {\bf w}_k \right\vert^2}{\sigma_c^2 + \sum^K_{j = 1, j \neq k}\left\vert {\bf h}_k^H {\bf w}_j \right\vert^2} \geq \gamma_k, \forall k. \label{eq:p3d}
		\end{align}
\end{subequations}

Obviously, problem (\ref{eq:p3}) is also intractable due to the fractional objective function \eqref{eq:p3a} and nonconvex constraints \eqref{eq:p3c} and \eqref{eq:p3d}. 
For handling the fractional objective function \eqref{eq:p3a}, with the introduced auxiliary optimization variables $\omega, t,\phi$, and $\zeta$, problem (\ref{eq:p3}) can be reformulated as
\begin{subequations} \label{eq:p3-1}
	\begin{align}
		\max_{ \{\mathbf{w}_k\}_{k=1}^{K}, \omega, \phi, \zeta}~~~~&\omega \label{eq:p3a1}\\
		\mathrm{s.t.}  ~~~~&\text{CRB}^{-1}(\theta)  \leq \frac{1}{t},  \label{eq:p3a11}  \\
		&	\frac{1}{\epsilon}\sum_{k=1}^K \|{\bf w}_k\|_2^2 + P_0 \leq \phi, t \geq \zeta^2, \label{eq:p3a12} \\
		& \omega \leq \frac{\zeta^2}{\phi},  \label{eq:p3a13} \\
		&\eqref{eq:p3b}, \eqref{eq:p3c}, \eqref{eq:p3d}. \notag 
	\end{align}%
\end{subequations}
The equivalence between \eqref{eq:p3-1} and \eqref{eq:p3} is obvious, since  constraints 
\eqref{eq:p3a1}, \eqref{eq:p3a11}, and \eqref{eq:p3a12} should be active at the optimal solution. We note that \eqref{eq:p3a11} share the same form with \eqref{eq:p1c}. Therefore, with Schur complement,  constraint \eqref{eq:p3a11} can be reformulated as
\begin{align}
	\begin{bmatrix}
	\mathcal{F}(\sum_{k=1}^K\mathbf{W}_k) - \frac{t \sigma_s^2}{2L \left\vert \alpha \right\vert^2}      & \sqrt{M}{\bf a}^{H}(\theta)\sum_{k=1}^K\mathbf{W}_k^{T}\dot{\bf a}(\theta)      \\
	\sqrt{M}\dot{\bf a}^{H}(\theta){\bf R}_{\bf x}^{T}{\bf a}(\theta)     & {\bf a}^{H}(\theta){\bf R}_{\bf x}^{T}{\bf a}(\theta) 
\end{bmatrix} \succeq \mathbf{0}, \label{eq:p3a2}
\end{align}
where  $\mathcal{F}(\sum_{k=1}^K\mathbf{W}_k) \triangleq M\dot{\bf a}^{H}(\theta)\sum_{k=1}^K\mathbf{W}_k^{T}\dot{\bf a}(\theta)+{\bf a}^{H}(\theta)\sum_{k=1}^K\mathbf{W}_k^{T}{\bf a}(\theta)\left\Vert\dot{\bf a}(\theta)\right\Vert^{2} $ and $\mathbf{W}_k = \mathbf{w}_k \mathbf{w}_k^H$.  Furthermore,  Lemma~\ref{lm:Www} presents an equivalent formulation of  the equality $\mathbf{W}_k = \mathbf{w}_k \mathbf{w}_k^H$ whose convex approximation has been given in \eqref{eq:rank1a} and \eqref{inequality_constraint_added}.
Then, for handling the fractional constraint \eqref{eq:p3d}, we introduce auxiliary variables $\left\lbrace \tau_k, \psi_k, \forall k\right\rbrace $ to reformulate \eqref{eq:p3d} as
\begin{subequations}
	\begin{align}
		&\tau^2_k / \psi_k \geq \gamma_k, \label{eq:sinr-sca-a} \\
		&\tau_k = \mathbf{h}_k^H \mathbf{w}_k,  \label{eq:eq:sinr-sca-b} \\
		&\psi_k \geq \sigma_c^2 + \sum^K_{j = 1, j \neq k}\left\vert {\bf h}_k^H {\bf w}_j \right\vert^2, \label{eq:sinr-sca-c}
	\end{align}
\end{subequations}
where \eqref{eq:eq:sinr-sca-b} and \eqref{eq:sinr-sca-c} are convex constraints. Then, problem \eqref{eq:p3} can be reformulated as
\begin{subequations} \label{eq:p3-3}
	\begin{align}
		\max_{\Theta}~~~&\omega \\
		\mathrm{s.t.}  ~~~ & \omega \leq \frac{\zeta^2}{\phi} , \gamma_k \leq \frac{\tau^2_k}{\psi_k} , \forall k \label{eq:p3-3a} \\
		&\eqref{matrix_inequality_constraint_added1}, \eqref{eq:p3b}, \eqref{eq:p3a12}, \eqref{eq:p3a2}, \eqref{eq:rank1a}, \eqref{inequality_constraint_added},\eqref{eq:eq:sinr-sca-b}, \eqref{eq:sinr-sca-c},  \label{eq:p3-3c}
	\end{align}
\end{subequations}
where $\Theta \triangleq \left\lbrace \{\mathbf{W}_k, \mathbf{w}_k\}_{k=1}^{K}, \omega, t,\phi, \zeta, \tau_k, \psi_k \right\rbrace $ denotes the set of  optimization variables. Obviously constraint \eqref{eq:p3-3c} is convex. Therefore, the challenge for handling problem \eqref{eq:p3-3} lies in the nonconvexity of constraint \eqref{eq:p3-3a}. To deal with this, we adopt the SCA techniques to establish a convex approximation of constraint \eqref{eq:p3-3a}. Since function $ \frac{\zeta^2}{\phi}$ is jointly convex with respect to $\zeta$ and $\phi$, its convex lower approximation can be established as
\begin{align}
	\frac{\zeta^2}{\phi} & \geq \frac{(\zeta^{(n)})^2}{\phi^{(n)}} +  \frac{2 \zeta^{(n)}}{\phi^{(n)}}   (\zeta  - \zeta^{(n)} )  - \left(  \frac{\zeta^{(n)}}{\phi^{(n)}} \right) ^2 (\phi  - \phi^{(n)} ) = \frac{2 \zeta^{(n)}}{\phi^{(n)}}   \zeta  - \left(  \frac{\zeta^{(n)}}{\phi^{(n)}} \right) ^2 \phi , \label{eq:ap1}
\end{align}
where $\zeta^{(n)}$ and $ \phi^{(n)}$ are the feasible points obtained at the $n$-th iteration of the SCA. Consequently, the inner convex approximation of $\omega \leq \frac{\zeta^2}{\phi}$ is 
\begin{equation}
\omega \leq \frac{2 \zeta^{(n)}}{\phi^{(n)}}   \zeta  - \left(  \frac{\zeta^{(n)}}{\phi^{(n)}} \right) ^2 \phi. \label{eq:sca1}
\end{equation}

Similarly, the inner convex approximation of $ \gamma_k \leq \frac{\tau^2_k}{\psi_k}, \forall k$ is
\begin{equation}
\gamma_k \leq \frac{2 \tau_k^{(n)}}{\psi_k^{(n)}}   \tau_k  - \left(  \frac{\tau_k^{(n)}}{\psi_k^{(n)}} \right) ^2 \psi_k , \forall k , \label{eq:sca2}
\end{equation}
where $\tau_k^{(n)}$ and $ \psi_k^{(n)}$ are the feasible points obtained at the $n$-th iteration.

Finally, a convex approximation of problem \eqref{eq:p3-3} is formulated as 
\begin{align}  \label{eq:p3-4}
		\max_{ \Theta}~~~&\omega \\
		\mathrm{s.t.}  ~~~ &  \eqref{eq:sca1},  \eqref{eq:sca2}, \eqref{eq:p3-3c}. \notag
	\end{align}

In this way, problem \eqref{eq:p3-4} can be solved with off-the-shelf numerical convex program solvers such as CVX Toolbox~\cite{grant2014cvx}. We summarize the proposed iterative method in Algorithm~\ref{alg2}, where its initial feasible solution can be obtained by following the penalty SCA method given in Remark 1.

\begin{algorithm}
	\caption{{\it: Proposed Iterative Algorithm for Handling \eqref{eq:p3}}}
	\label{alg2} 
	\begin{algorithmic}
		\STATE Initialize $n = 0$, $\Theta^{(0)} \in \mathcal{S}$.
		\REPEAT 
		\STATE $n \leftarrow n+1$;
        \STATE $\tilde{\mathbf{w}}_k^{(n)} \leftarrow \mathbf{w}_k^{(n-1)}$
		\STATE Solve problem \eqref{eq:p3-4} with  $\Theta^{(n-1)}$ and obtain the optimal value $\Theta^{(\ast)}$ ;
		\STATE$ \Theta^{(n)} \leftarrow \Theta^{(\ast)}$ ;
		\UNTIL Convergence   
	\end{algorithmic}
\end{algorithm}

In the following, we analyze the convergence of Algorithm~\ref{alg2}. We can note that in the iterative procedure of Algorithm \ref{alg2}, $\Theta^{(n-1)}$ is always feasible in problem \eqref{eq:p3-4} at $n$-th iteration  owing to the adopted first-order Taylor approximation. We note that \eqref{eq:p3-4} can be optimally solved and the optimal value of its objective function serves as a lower bound on that of \eqref{eq:p3-3}.
Therefore, it can be guaranteed that the optimal value of \eqref{eq:p3-3} at $n$-th iteration $n$, denoted as $p_\ast^{(n)}$, always satisfies $p_\ast^{(n)} \geq p_\ast^{(n-1)}$. Therefore, Algorithm \ref{alg2} produces a non-decreasing objective function of problem \eqref{eq:p3-3}.
Similar to Algorithm \ref{alg1}, the computational complexity of Algorithm \ref{alg2} is  $\mathcal{O}( \sqrt{(2M +1)(K+1)} M^6 K^3 I_\mathrm{iter} \ln(1/\epsilon_0) )$.

\subsection{Extened Target Case} \label{sec:sensing-e}

For the case of the extended target, following the discussion in Section~\ref{sec:com-extend}, we choose $\mathbf{A}$  as the parameter to be estimated and adopt the formulation of CRB in \eqref{eq:CRB_extended}.
Then, we have the sensing-centric EE for sensing an extended target as 
\begin{align}
    \mathrm{EE}_\mathrm{S} = \frac{ \left( \frac{\sigma_s^2 M}{L} \operatorname{tr} \left({{\mathbf {R_x}}^{-1}} \right) \right)^{-1}}{ L \left( \frac{1}{\epsilon}  \operatorname{tr}(\mathbf {R_x})  + P_0 \right) } =  \frac{ \left(  \operatorname{tr} \left({\mathbf {R_x}^{ - 1}} \right) \right)^{-1}}{\sigma_s^2 M \left( \frac{1}{\epsilon}  \operatorname{tr}(\mathbf {R_x}) + P_0 \right) },
\end{align}
where ${{\mathbf {R}}_\mathbf{X}} = \mathbf{W} \mathbf{W}^H +  \tilde{\mathbf{W}} \tilde{\mathbf{W}}^H = \sum_{k=1}^K \mathbf{w}_k \mathbf{w}_k^H + \mathbf{R}_{\tilde{\mathbf{W}}}$. Then, we formulate the problem as
\begin{subequations} \label{eq:p4}
	\begin{align}
		\max_{ \{\mathbf{w}_k\}_{k=1}^{K},\mathbf{R}_{\tilde{\mathbf{W}}}}~~~& \frac{ \left(  \operatorname{tr} \left({\mathbf {R_x}^{ - 1}} \right) \right)^{-1}}{\sigma_s^2 M \left( \frac{1}{\epsilon}  \operatorname{tr}(\mathbf {R_x}) + P_0 \right) } \label{eq:p4a} \\ 
		\mathrm{s.t.}  ~~~		&\operatorname{tr}(\mathbf {R_x}) \leq P_{\text{max}}, \label{eq:p4b}  \\ 
		&\frac{\sigma_s^2 M}{N} \operatorname{tr} \left({{\mathbf {R_x}}^{ - 1}} \right)  \leq \phi , \label{eq:p4c} \\ 
		& \tilde{\text{SINR}_k} \geq \gamma_k, \label{eq:p4d}
	\end{align}
\end{subequations}
where $ \tilde{\text{SINR}_k}$ is given in \eqref{eq:sinr-extended} and can be recast as a convex form in \eqref{eq:p2convexd} by letting $\mathbf{W}_k = \mathbf{w}_k \mathbf{w}_k^H$.
We notice that in \eqref{eq:p4a}, the numerator is the reciprocal of a convex function and the denominator is strictly positive and convex. To handle its nonconvexity, we introduce auxiliary optimization variables $p_e,q_e$ and equivalently transform the problem into 
\begin{subequations}
\begin{align} \label{eq:p41}
	\max_{ \{\mathbf{w}_k\}_{k=1}^{K},\mathbf{R}_{\tilde{\mathbf{W}}}, q_e, p_e}~~& \frac{1}{p_e q_e}  \\ 
	\mathrm{s.t.} ~~~~~~ &p_e \geq  \sigma_s^2 M \left( \frac{1}{\epsilon}  \operatorname{tr}(\mathbf {R_x}) + P_0 \right), q_e \geq \operatorname{tr} \left({{\mathbf {R_x}}^{ - 1}} \right), \label{eq:p42b} \\
	& \eqref{eq:p4b}, \eqref{eq:p4c},\eqref{eq:p4d}. \label{eq:p42c}
\end{align}
\end{subequations}
Then, the problem can be further transformed into its equivalent form as
\begin{align}
		\min_{ \{\mathbf{W}_k\}_{k=1}^{K},\mathbf{R}_{\tilde{\mathbf{W}}}, q_e, p_e}~~& \ln(p_e) + \ln(q_e) ~~~~ \mathrm{s.t.}~~\eqref{eq:p42b},~ \eqref{eq:p42c}, \label{eq:p5a2} 
	\end{align}
where the objective function is still not convex, but can be approximated based on the first order Taylor series expansion given by
\begin{align}
\ln(p_e) + \ln(q_e) \leq \ln \left( p^{(n)}_e \right)  + \ln \left( q_e^{(n)}\right)  + \frac{1}{p_e^{(n)}} \left( p_e-p_e^{(n)} \right)  + \frac{1}{q^{(n)}_e} \left( q_e-q^{(n)}_e\right) ,
\end{align}
where $p_e^{(n)}$ and $q_e^{(n)}$ are the feasible solutions obtained at the $n$-th iteration. Following the techniques detailed in Section~\ref{sec:com-extend}, a convex approximation of problem \eqref{eq:p41} at the $n$-th iteration can be established as
\begin{subequations}
\begin{align} \label{eq:p4convex}
	\min_{ \{\mathbf{W}_k\}_{k=1}^{K}, \mathbf{R}_{\tilde{\mathbf{W}}}, q_e, p_e}~& \ln(p^{(n)}_e) + \ln(q_e^{(n)}) + \frac{1}{p_e^{(n)}} (p_e-p_e^{(n)}) + \frac{1}{q^{(n)}_e} (q_e-q^{(n)}_e)\\ 
	\mathrm{s.t.} ~	~	& 	\eqref{eq:p2convexb}, \eqref{eq:p2convexc},\eqref{eq:p2convexd},\eqref{eq:p2convexe}, \eqref{eq:p42b}. 
\end{align}
\end{subequations}
The computational complexity is $\mathcal{O}( \sqrt{MK+M+K+1} M^6 K^3 I_\mathrm{iter} \ln(1/\epsilon_0) )$ for an $\epsilon_0$-optimal solution.

\begin{Th}
Based on the optimal solution of \eqref{eq:p4convex}, denoted as $\{\mathbf{W}_k^\ast, \mathbf{R}^\ast_{\tilde{\mathbf{W}}} \}$, the optimal rank-1 solutions can always be reconstructed.
\end{Th}

\begin{IEEEproof}
The proof can be achieved by following the proof of Theorem 2 and the details are omitted for brevity.
\end{IEEEproof}




\section{Approximate Pareto Boundary of Energy-Efficient ISAC Systems}

In this section, we aim to investigate the Pareto boundary of the achievable EE performance region built on the communication-centric EE and the sensing-centric EE. 
Considering the point-like target case, we follow~\cite{gao2023cooperativepareto} to formulate the search of the Pareto boundary  as a constrained optimization problem that maximizes the communication-centric EE under the sensing-centric EE constraint. It is worth noting that the proposed algorithm can be adapted to the extended target case directly. Now, we aim to solve
\begin{subequations}\label{addproblem}
	\begin{align}
		\max_{ \{\mathbf{w}_k\}_{k=1}^{K}}~~&\frac{\sum_{k=1}^K  \log_2 \left( 1+\left\vert {\bf h}_k^H {\bf w}_k \right\vert^2 \bigg/ \left( \sigma_c^2 + \mathop{\sum}_{k \in \mathcal{K} \atop j \neq k}\left\vert {\bf h}_k^H {\bf w}_j \right\vert^2\right)  \right) }{\frac{1}{\epsilon}\sum_k \|{\bf w}_k\|_2^2 + P_0} \label{eq:p5a}\\
		\mathrm{s.t.}  ~~&\frac{\text{CRB}^{-1}(\theta)}{ L \left(  \frac{1}{\epsilon}\sum_{k=1}^K  \|{\bf w}_k\|_2^2 + P_0 \right) } \geq \mathcal{E}, \label{eq:p5ees}\\
        &\sum_k \|{\bf w}_k \|_2^2 \leq P_\text{max}, \label{eq:p5b}
	\end{align}
\label{eq:p5}%
\end{subequations}
where $\mathcal{E}$ denotes the required minimum sensing-centric EE threshold. 
Obviously, problem \eqref{addproblem} is a nonconvex fractional program, which is challenging to solve directly. 
To handle  fractional objective function \eqref{eq:p5a} and  nonconvex constraint \eqref{eq:p5ees}, we follow~\cite{gao2023cooperativepareto} to find the approximate optimal Pareto boundary for characterizing the  tradeoff between the communication-centric EE and sensing-centric EE. 

In particular, we first apply the Dinkelbach algorithm to reformulate  fractional function~\eqref{eq:p5a} as
\begin{align}
 &\max_{\lambda}~{\sum_{k=1}^{K}  \log_2 \left( 1+ \frac{ \left\vert {\bf h}_k^H {\bf w}_k \right\vert^2 } {{B}_k(\mathbf{W}_k)}  \right)  }  -  \lambda  \left(  {\frac{1}{\epsilon} \sum_{k=1}^{K} \text{tr}({\bf W}_k)+ P_0 }\right) \nonumber\\
 &\mathrm{s.t.}~ \eqref{eq:rank1a},~\eqref{eq:rank1b},
 \label{eq:p5ob1}
\end{align} 
where $B_k(\mathbf{W}_k) = \sum^K_{j=1, j \neq k} \text{tr}(\mathbf{Q}_k \mathbf{W}_j) + \sigma_c^2$.

Furthermore, by introducing auxiliary variables $b_k$, $k=1,\ldots,K$, the intractable fractional terms in (\ref{eq:p5ob1}) can be equivalently formulated as
\begin{align}
	&  {\sum_{k=1}^{K}\log_2 \left( 1+ \frac{ \left\vert {\bf h}_k^H {\bf w}_k \right\vert^2 } {{B}_k(\mathbf{W}_k)}  \right)  }  = \max_{b_k}~ \left( \sum_{k=1}^{K} \log_2 (1+ b_k)  - \sum_{k=1}^{K} b_k + \sum_{k=1}^{K} \frac{(1+b_k)\left\vert {\bf h}_k^H {\bf w}_k \right\vert^2 }{ {B}_k(\mathbf{W}_k)} \right),  \label{eq:p5ob2}
\end{align} 
which has an analytical solution $b_k = \frac{\left\vert {\bf h}_k^H {\bf w}_k \right\vert^2}{{B}_k(\mathbf{W}_k)}$.
Finally, by applying the quadratic transform~\cite[Theorem 1]{shen2018fractional}, problem \eqref{eq:p5} can be reformulated as 
\begin{subequations}
	\begin{align}
		\max_{ \{\mathbf{w}_k \mathbf{W}_k, b_k, t_k\}_{k=1}^{K}, \lambda}~~& \sum_k \left(  \log_2 (1+ b_k)  - b_k + 2t_k \sqrt{(1+b_k)} \operatorname{Re}(\mathbf{w}_k^H \mathbf{h}_k) - t_k^2  {B}_k(\mathbf{W}_k)  \right) \notag\\
  & -  \lambda \left( {\frac{1}{\epsilon} \sum_{k=1}^{K} \text{tr}({\bf W}_k)+ P_0}\right) \label{eq:p5convexa1} \\
		\mathrm{s.t.}  ~~~~&  \eqref{eq:rank1a},  \eqref{eq:rank1b},\eqref{eq:p5ees},\eqref{eq:p5b}.
  \label{eq:p5constraints}
	\end{align}
	\label{eq:p5convex1}%
\end{subequations}
The convex approximation of nonconvex constraint \eqref{eq:rank1b} is constraint \eqref{inequality_constraint_added},  as mentioned in ~Section \ref{sec-comm-point}. For handling nonconvex constraint \eqref{eq:p5ees}, 
we introduce an auxiliary variable $\tilde{\mathcal{E}}$ and employ the Schur complement to obtain the convex approximation of problem  \eqref{eq:p5} given by
\begin{subequations}
	\begin{align}
		\max_{ \{\mathbf{w}_k \mathbf{W}_k, b_k, t_k\}_{k=1}^{K}, \lambda}~~& \sum_k \left(  \log_2 (1+ b_k)  - b_k + 2t_k \sqrt{(1+b_k)} \operatorname{Re}(\mathbf{w}_k^H \mathbf{h}_k) \!- t_k^2 {B}_k(\mathbf{W}_k)   \right)    \notag \\
		& -  \lambda \left( {\frac{1}{\epsilon} \sum_{k=1}^{K} \text{tr}({\bf W}_k)+ P_0}\right) \label{eq:p5convexa} \\
		\mathrm{s.t.}  ~~~~& \begin{bmatrix}
	\mathcal{F}(\sum_{k=1}^K\mathbf{W}_k) - \frac{\tilde{\mathcal{E}} \sigma_s^2}{2L \left\vert \alpha \right\vert^2}      & \sqrt{M}{\bf a}^{H}(\theta)\sum_{k=1}^K\mathbf{W}_k^{T}\dot{\bf a}(\theta)      \\
	\sqrt{M}\dot{\bf a}^{H}(\theta){\bf R}_{\bf x}^{T}{\bf a}(\theta)     & {\bf a}^{H}(\theta){\bf R}_{\bf x}^{T}{\bf a}(\theta) 
\end{bmatrix} \succeq \mathbf{0} ,\\
&\tilde{\mathcal{E}} \geq \mathcal{E} N \left({\frac{1}{\epsilon} \sum_{k=1}^{K} \text{tr}({\bf W}_k)+ P_0} \right),\\
& \eqref{eq:rank1a}, \eqref{inequality_constraint_added}, \eqref{eq:p5b}.
	\end{align}
	\label{eq:p5convex}%
\end{subequations}
\eqref{eq:p5convex} is convex whose optimum can be obtained by the interior point method. Therefore, an efficient solution of problem  \eqref{eq:p5} can be obtained by solving a sequence of problem \eqref{eq:p5convex}. Algorithm \ref{alg3} summarizes the iterative algorithm, where $\breve{f}_1(\mathbf{w}_k, \mathbf{W}_k)\! = \frac{\beta}{\overline{\mathcal{R}} } \sum_{k=1}^K\!\! \left( \log_2 (1\!+\! b_k) \! \!- \! b_k\! + \!2t_k \sqrt{(1+b_k)} \operatorname{Re}(\mathbf{w}_k^H \mathbf{h}_k)\right.$ $ \! - \!t_k^2 {B}_k(\mathbf{W}_k) \Big)  + (1-\beta) \frac{\tilde{\phi}}{L \overline{\mathcal{C}}} $, $\breve{f}_2(\mathbf{W}_k) = \lambda \left( {\frac{1}{\epsilon} \sum_{k=1}^{K} \text{tr}({\bf W}_k)+ P_0}\right) $.

\begin{algorithm} 
	\caption{{\it: The Proposed Algorithm for Handling Problem \eqref{eq:p5}}}
	\label{alg3} 
	\begin{algorithmic}
		\STATE Initialize $i = 0$, $\delta > 0$, $ \{ \mathbf{w}^{(0)}_k, \mathbf{W}^{(0)}_k\}_{k=1}^{K}$ to a feasible value;
		\REPEAT 
		\STATE $i \leftarrow i + 1$ ;
        \STATE $\tilde{\mathbf{w}}_k^{(i)} \leftarrow \mathbf{w}_k^{(i-1)}$
		\STATE Update $\lambda$ by $\lambda^{(i)} = \frac{\breve{f}_1(\mathbf{w}^{(i-1)},\mathbf{W}^{(i-1)})}{\breve{f}_2(\mathbf{W}^{(i-1)})}$;
		\STATE Update $b_k$ by $ b^{(i)}_k =  \frac{ |{\mathbf{w}_k^{(i-1)}}^H \mathbf{h}_k)|^2 } {{B}_k(\mathbf{W}_k^{(i-1)})} $;
		\STATE Update $t_k$ by $ t^{(i)}_k = \frac{\sqrt{(1+b_k)}\operatorname{Re}({\mathbf{w}_k^{(i-1)}}^H \mathbf{h}_k
}{B_k(\mathbf{W}_k^{(i-1)})}$;
		\STATE Solve problem \eqref{eq:p5convex} to obtain the optimal $\mathbf{w}_k^{(i)}, \mathbf{W}_k^{(i)}$ ;
		\UNTIL $\breve{f}_1(\mathbf{w}^{(i)},\mathbf{W}^{(i)})-\lambda^{(i)}{\breve{f}_2(\mathbf{W}^{(i)})}$ is less than $\delta$.   
	\end{algorithmic}
\end{algorithm}

\section{Numerical Results}
In this section, we provide simulation results of the proposed energy-efficient waveform design. Numerical analysis is presented to evaluate the performance of communication-centric EE ($\text{EE}_\text{C}$), sensing-centric EE ($\text{EE}_\text{S}$), and their approximate Pareto boundary.
Unless stated otherwise, we consider a dual-functional BS equipped $N = 20$ receiving antennas, with the frame length $N$ set to $30$. The maximum transmission power $P_{\text{max}}$ is set to $30$ dBm with the power amplifier efficiency $\epsilon = 0.35$. The circuit power consumption is set to $P_0 = 33 $ dBm. For the target estimation of radar, the target angle is $\theta = 90 ^\circ$. 
\subsection{$\text{EE}_\text{C}$ Optimization}

\begin{figure}
     \centering
     \begin{subfigure}[b]{0.4\textwidth}
         \centering
         \includegraphics[width=0.9\textwidth]{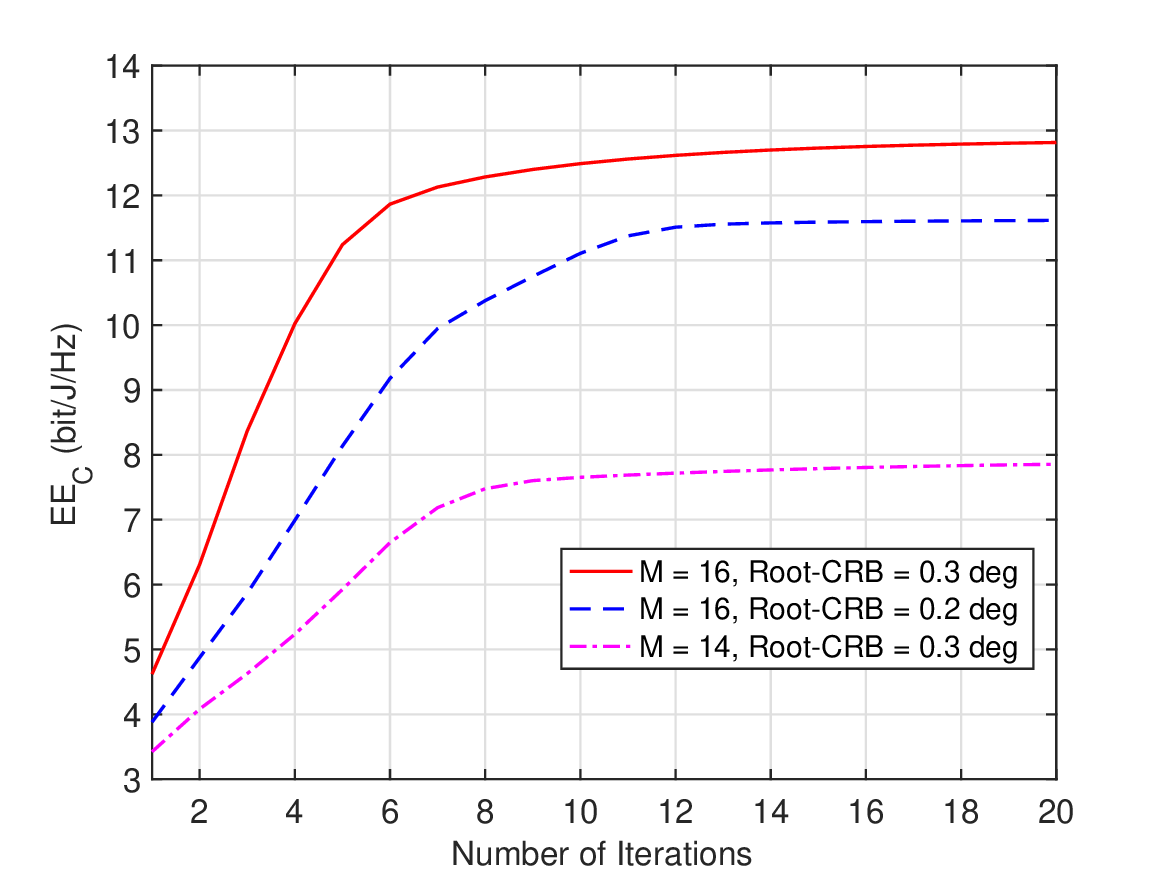}
         \caption{}
         \label{fig:ee-iter}
     \end{subfigure}
     \begin{subfigure}[b]{0.4\textwidth}
         \centering
         \includegraphics[width=0.9\textwidth]{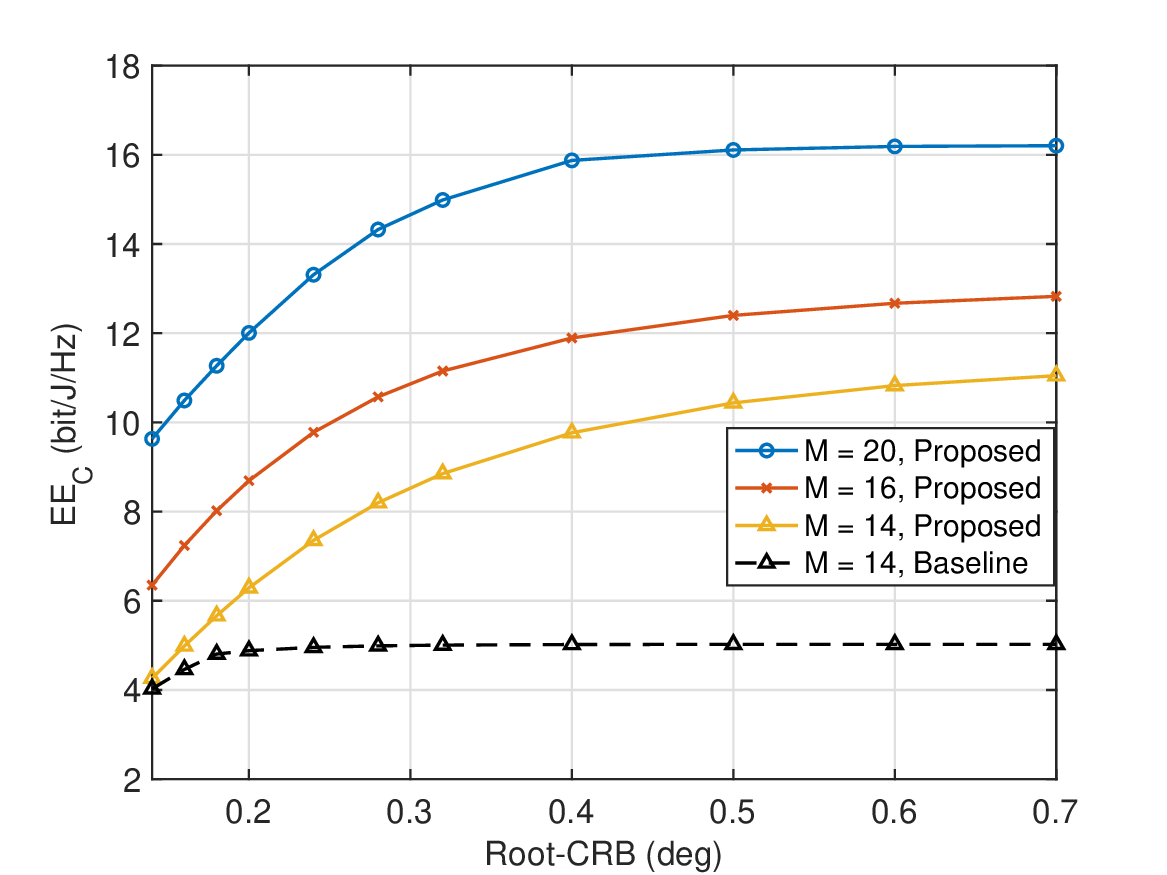}
           \caption{}
         \label{fig:ee-crb}
     \end{subfigure}
     \hfill
        \caption{(a) $\text{EE}_\text{C}$ versus the number of iterations with $\gamma_k = 10$ dB, $K = 3$ for the point-like target case; (b) $\text{EE}_\text{C}$ versus different root-CRB thresholds with  $\gamma_k = 10$ dB, $K = 2$ for the point-like target case.}
\end{figure}

We first examine the performance of Algorithm~\ref{alg1} for maximizing $\text{EE}_\text{C}$ considering the existence of a point-like target. The convergence rate of Algorithm 1 is given in Fig.~\ref{fig:ee-iter}. Obviously, it enjoys a fast convergence rate, whose objective function value converges within 12 iterations on average. 
Furthermore, the convergence rate of  Algorithm 1 is almost the same for
different system parameters, e.g., different $M$ and CRB constraints, which confirms the scalability  of  Algorithm 1.

Fig.~\ref{fig:ee-crb} investigates the $\text{EE}_\text{C}$ performance versus the root-CRB threshold for different $M$. The $\text{EE}_\text{C}$ increases with the increasing Root-CRB threshold, indicating that $\text{EE}_\text{C}$ can achieve a higher level when the sensing performance requirement is less stringent. Indeed, increasing the number of antennas can improve $\text{EE}_\text{C}$, since more spatial degrees-of-freedom can be utilized for designing an efficient ISAC waveform. On the other hand, the baseline scheme only maximizes the communication sum rate under the same constraints of problem \eqref{original_problem}. 
Obviously, the $\text{EE}_\text{C}$ of the baseline scheme is unsatisfying, since it only considers the spectral efficiency maximization instead of the  $\text{EE}_\text{C}$ maximization. In such a case, the baseline scheme encourages the ISAC BS to adopt as much power as possible for increasing the communication sum rate. 


\begin{figure}
     \centering
     \begin{subfigure}[b]{0.4\textwidth}
         \centering
         \includegraphics[width=0.9\linewidth]{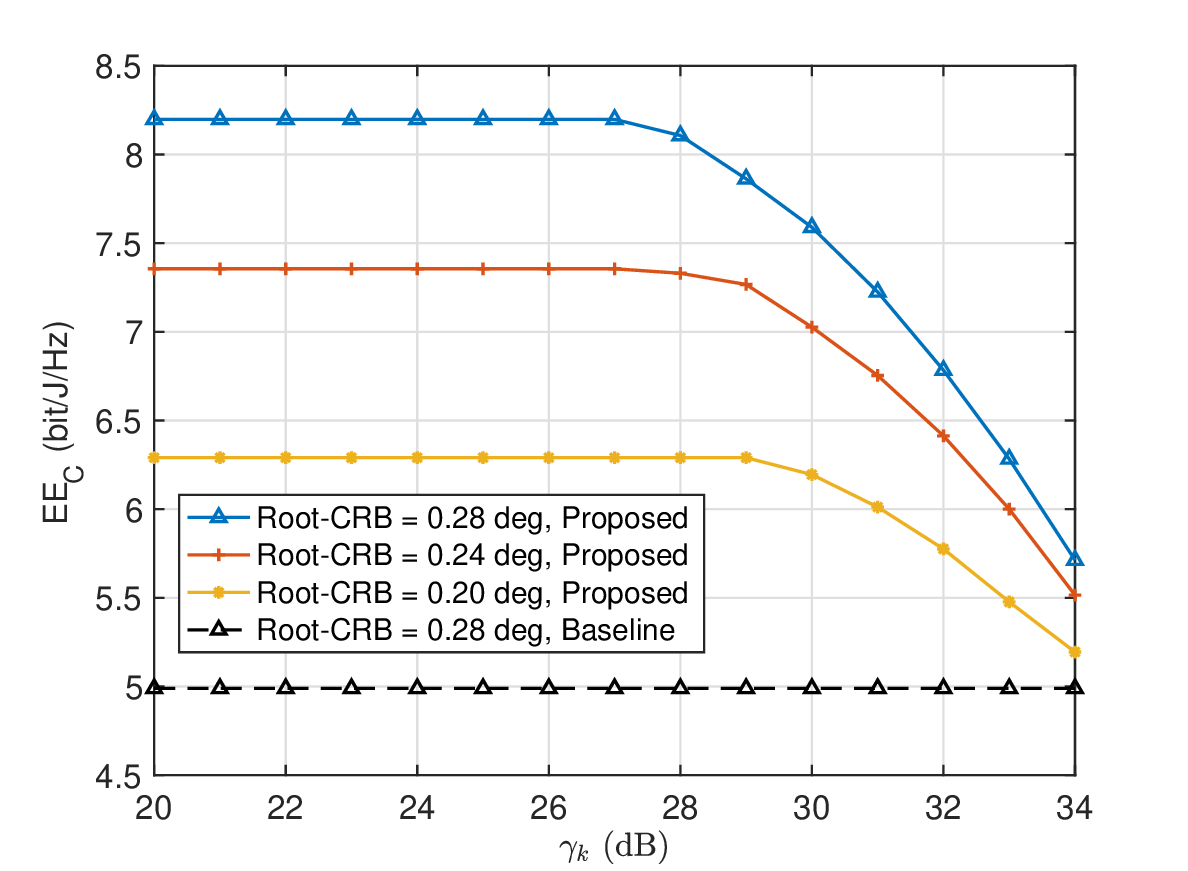}
	\caption{}
	\label{fig:ee-MSE}
     \end{subfigure}
     \begin{subfigure}[b]{0.4\textwidth}
         \centering
         \includegraphics[width=0.9\linewidth]{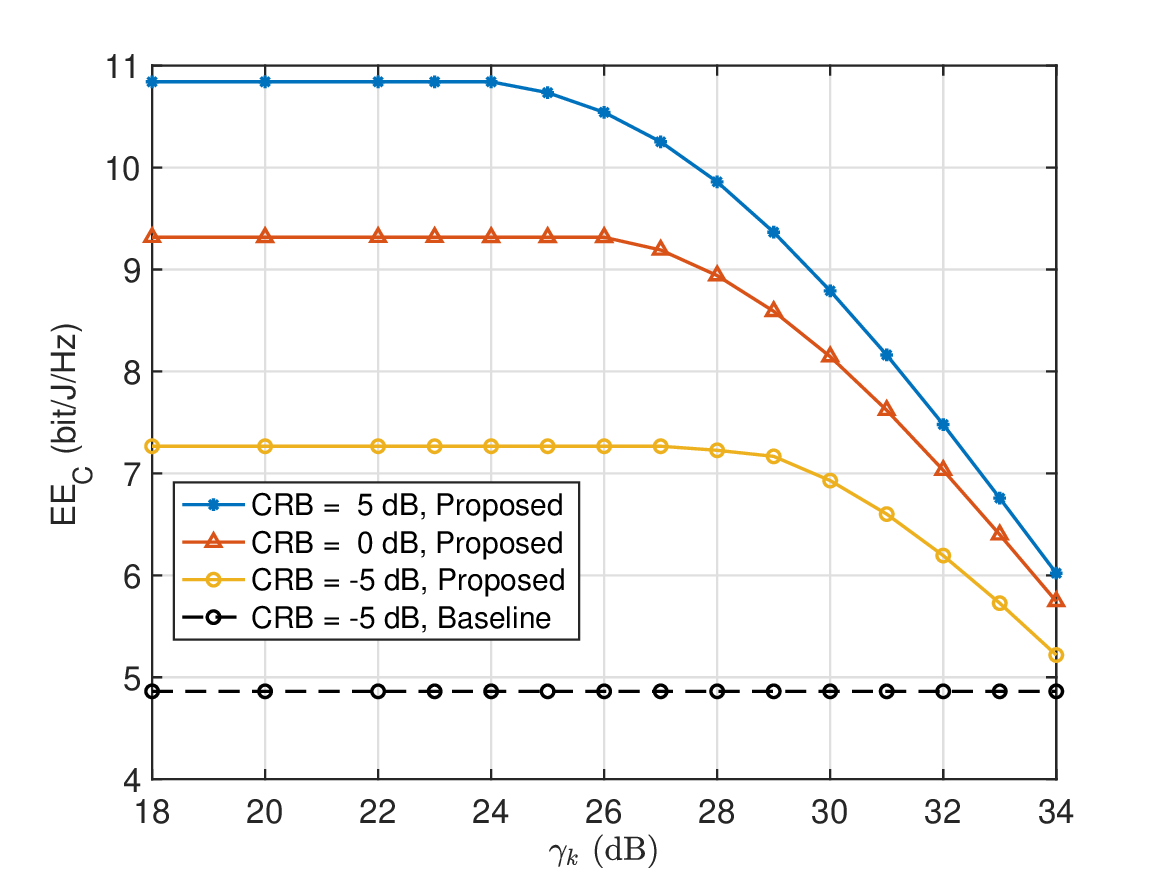}
	\caption{}
	\label{fig:ee-antenna}
     \end{subfigure}
     \hfill
        \caption{$\text{EE}_\text{C}$ versus different SINR requirements $\gamma_k$. (a) The point-like target case with $M=14, K=2$; (b) The extended target case with $M=14, K=2$.}
\end{figure}

Fig. \ref{fig:ee-MSE} and Fig. \ref{fig:ee-antenna} plot the $\text{EE}_\text{C}$ of the point-like target and extended target  with the increasing SINR constraint of multiple users, $\gamma_k$, respectively. With the increasing $\gamma_k$, $\text{EE}_\text{C}$ first remains unchanged and then decreases due to the shrunken feasible region. Therefore, increasing the downlink communication rate does not necessarily improve $\text{EE}_\text{C}$. Furthermore, with the increasing root-CRB, the  $\text{EE}_\text{C}$ decreases, since more power is allocated to radar sensing due to the increasing sensing requirements. A similar trend can also be found in Fig. \ref{fig:ee-antenna} for the increasing CRB in the extended target case. 




\subsection{$\text{EE}_\text{S}$ Optimization}


\begin{figure}
     \centering
     \begin{subfigure}[b]{0.4\textwidth}
         \centering
         \includegraphics[width=0.9\linewidth]{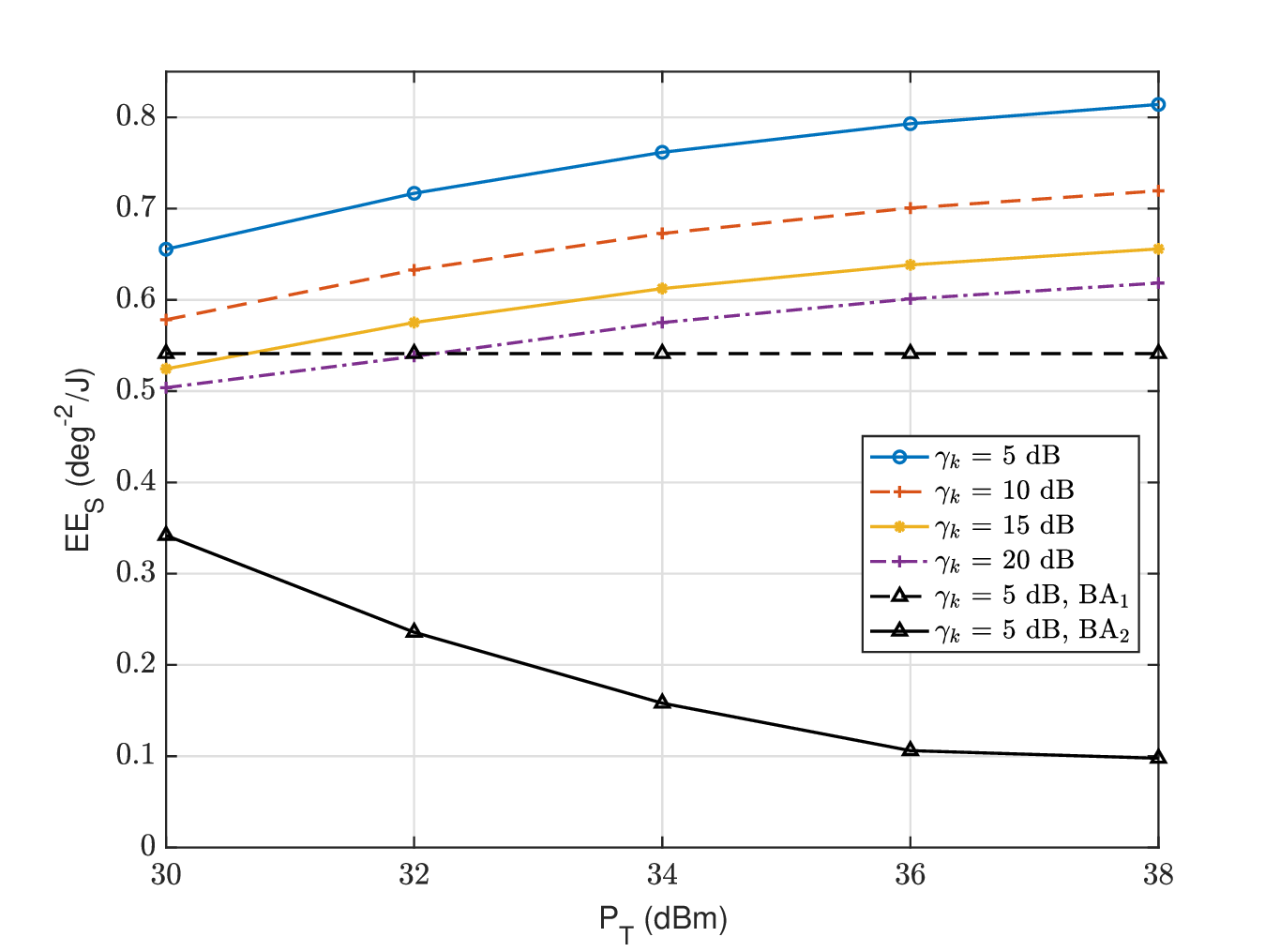}
	\caption{}
	\label{fig:ees-pt-sinr}
     \end{subfigure}
     \begin{subfigure}[b]{0.4\textwidth}
         \centering
         \includegraphics[width=0.9\linewidth]{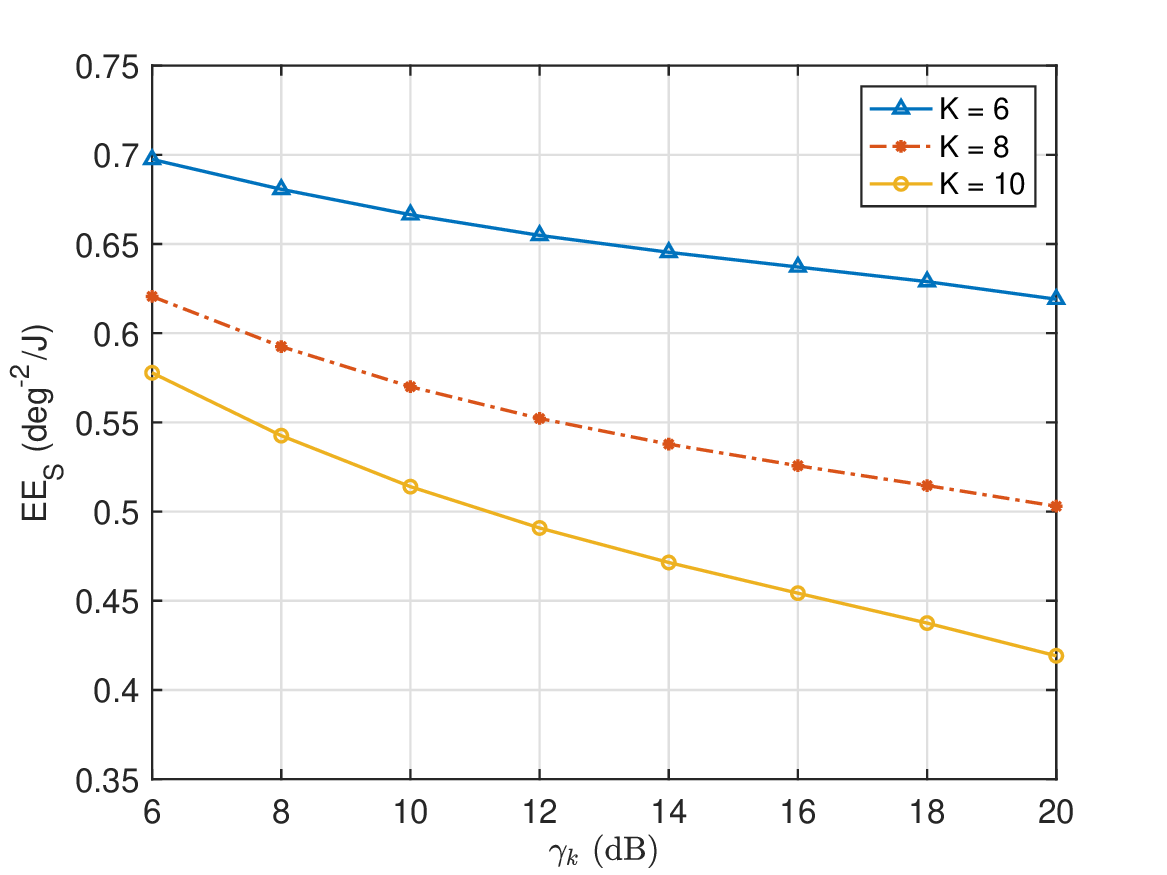}
	\caption{}
	\label{fig:ees-sinr-k}
     \end{subfigure}
     \hfill
        \caption{(a) $\text{EE}_\text{S}$ for the point-like target case versus the maximum transmission power $P_T$, compared with the baselines, with $M = 16, \rho = 0.15$ deg, $K=8$. (b) $\text{EE}_\text{S}$ for the point-like target case versus the SINR requirements, under different numbers of users, with $M = 16, \rho = 0.15$ deg.}
        \vspace{-1em}
\end{figure}



In this subsection, we investigate the performance of $\text{EE}_\text{S}$ optimization for both the point-like target sensing and extended target cases.  In Fig.~\ref{fig:ees-pt-sinr}, we first consider the point-like target to show the $\text{EE}_\text{S}$ versus the increasing power budget, for different SINR levels. As expected, $\text{EE}_\text{S}$ increases with the increasing $P_T$, since the increasing power improves the estimation accuracy and increases $\text{EE}_\text{S}$. Besides, lowering the SINR requirement also improves
$\text{EE}_\text{S}$, since relaxing the SINR constraint enlarges the feasible region and improves $\text{EE}_\text{S}$.
For demonstrating the performance gain obtained by our proposed Algorithm 3, 
we perform the performance comparison with two other baselines, namely $\text{BA}_1$ and $\text{BA}_2$. In particular, $\text{BA}_1$ aims to minimize the transmission power while $\text{BA}_2$ aims to maximize the communication sum rate under the same constraints as our proposed method ($\gamma_k$ = 5 dB, the root-CRB threshold is set to 0.15 deg, $P_{\text{max}} = 30 $ dBm). The results indicate that $\text{EE}_\text{S}$ of $\text{BA}_1$ is significantly low due to the insufficient  power for improving the CRB performance. Additionally, $\text{EE}_\text{S}$ of $\text{BA}_2$ is also inferior to the proposed method and exhibits a further decline as the transmission power increases, since most of the power is utilized for maximizing the sum rate instead of sensing target.

Fig.~\ref{fig:ees-sinr-k} further demonstrates the $\text{EE}_\text{S}$ versus the SINR requirement, where the root-CRB threshold is set to 0.15 deg. It can be observed  that $\text{EE}_\text{S}$ decreases as the increasing SINR and the number of communication users since the increasing communication requirements deteriorates the sensing performance.


\begin{figure}
     \centering
     \begin{subfigure}[b]{0.4\textwidth}
         \centering
         \includegraphics[width=0.9\linewidth]{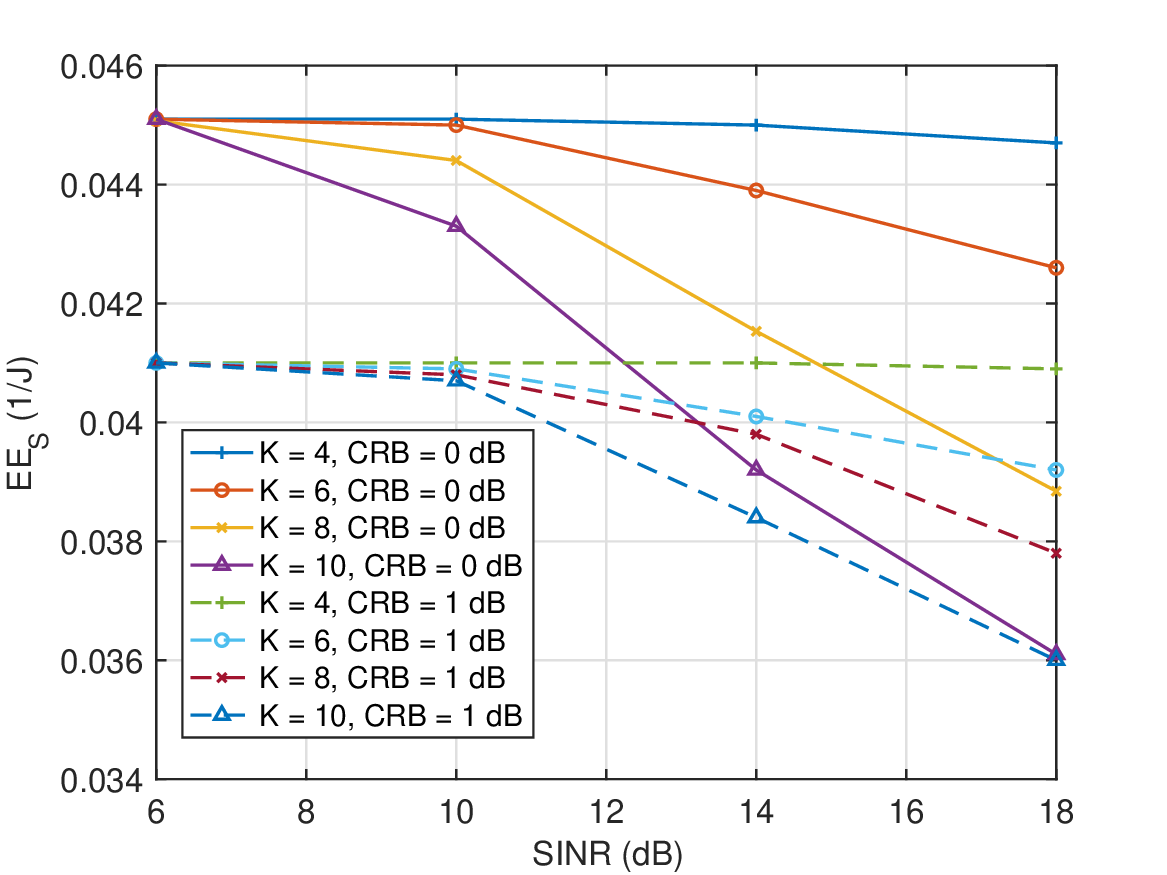}
	\caption{}
	\label{fig:EES_SINR_MSE}
     \end{subfigure}
     \begin{subfigure}[b]{0.4\textwidth}
         \centering
         \includegraphics[width=0.9\linewidth]{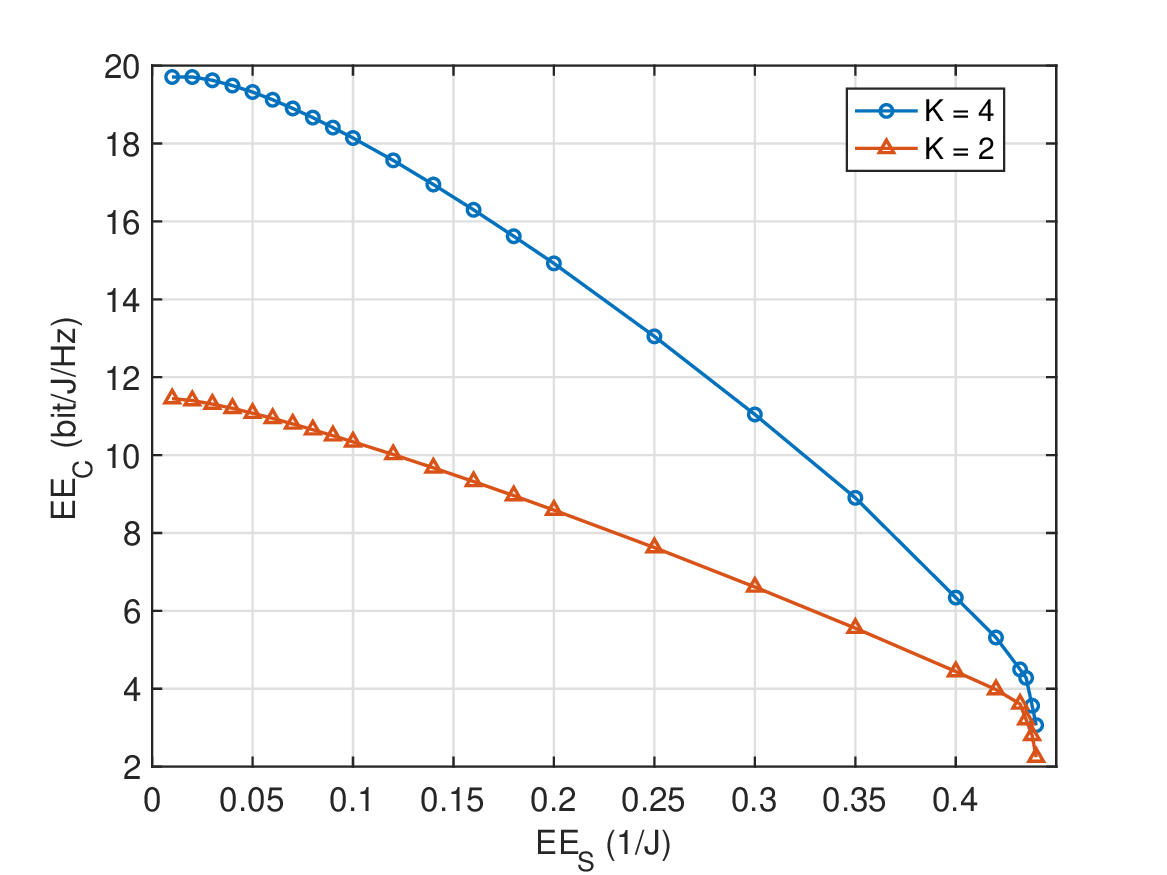}
	\caption{}
	\label{fig:pareto}
     \end{subfigure}
     \hfill
        \caption{(a) $\text{EE}_\text{S}$ in the scenario of sensing an extended target versus the SINR requirement, under different CRB constraints and numbers of users. (b) The Pareto boundary of energy-efficient ISAC for different numbers of communication users with $M = 14$ and $P_{\max}=30$ dBm.}

        \vspace{-1em}
\end{figure}

As for the scenario of sensing an extended target, Fig.~\ref{fig:EES_SINR_MSE} shows the $\text{EE}_\text{S}$ versus communication SINR under different numbers of users and different CRB. 
It is worth noting that  the performance metric for the extended target sensing $\text{EE}_\text{S}$ is different from the point-like target case.
Similar to the scenario of sensing a point-like target, $\text{EE}_\text{S}$ decreases with the increasing requirements of communication SINR, especially when the number of users is larger. Besides, increasing  CRB requirements improves $\text{EE}_\text{S}$, due to the improved estimation performance. 
 
%
\subsection{Approximate Pareto Boundary of Energy-Efficient ISAC.}

Fig.~\ref{fig:pareto} plots the approximate Pareto boundary of energy-efficient ISAC, which demonstrates the tradeoff between $\text{EE}_\text{C}$ and $\text{EE}_\text{S}$.  With the more stringent $\text{EE}_\text{S}$ constraint, the $\text{EE}_\text{C}$ decreases. 
In particular, when the required minimum sensing-centric EE threshold $\mathcal{E}$ is small, strengthening the requirement of  $\text{EE}_\text{S}$ only affects $\text{EE}_\text{C}$ mildly.
However, when the required $\text{EE}_\text{S}$ beyonds a certain threshold, increasing $\text{EE}_\text{S}$ constraint will bring a sharp decline in $\text{EE}_\text{C}$.
This phenomenon shows that there is a non-trivial tradeoff between $\text{EE}_\text{S}$ and $\text{EE}_\text{C}$, which should be given serious consideration.
Besides, we can find that the area spanned by the Pareto boundary is sensitive to the  number of communication users, $K$, since the increasing number of served communication users consumes the available spatial degrees of freedom which cannot compensate for the performance loss due to the increasingly stringent $\text{EE}_\text{S}$ constraint. 
Therefore, it is more challenging to balance  $\text{EE}_\text{S}$ and $\text{EE}_\text{C}$ for a large $K$.
On the other hand, after the required $\text{EE}_\text{S}$ surpasses some threshold, $\text{EE}_\text{C}$ decreases sharply. This is because most of the available resources are allocated for satisfying the stringent EE$_\text{s}$ constraint, such that the remaining resources are insufficient for guaranteeing the $\text{EE}_\text{C}$ performance.


\section{Conclusion}

In this paper, we addressed the problem of maximizing energy efficiency for MIMO ISAC systems. We first studied the communication-centric EE adopting the conventional definition of EE in both the point-like target and extended target cases. We reformulated the objective function using the quadratic-transform-Dinkelbach method and solved the sub-problem by leveraging the Schur complement and semi-relaxation techniques. In the second part, we introduced a novel performance metric for measuring sensing-centric EE. We iteratively approximated the objective function as a convex program exploiting SCA to address this problem. Finally, we investigated the tradeoff between the two EE metrics and provided an effective solution. Numerical results showed an improvement compared to the benchmark on both communication-centric EE and sensing-centric EE performance, and we also demonstrated the tradeoff between communication-centric and sensing-centric EE.


{\appendices
\section*{Appendix A}
First, we provide the matrix inequality 
	$\mathbf{W}_k \succeq  \mathbf{w}_k \mathbf{w}_k^H, \label{eq:Www3}$
which satisfies either of the following cases:

\textbf{Case I}: $\mathbf{W}_k \succ \mathbf{w}_k \mathbf{w}_k^H$. Then, we have $\operatorname{tr}(\mathbf{W}_k) ~ > ~ \operatorname{tr}(\mathbf{w}_k \mathbf{w}^H_k)$.

\textbf{Case II}: $\mathbf{W}_k =  \mathbf{w}_k \mathbf{w}_k^H$. In this case, we have $\operatorname{tr}(\mathbf{W}_k) = \operatorname{tr}(\mathbf{w}_k \mathbf{w}^H_k)$.

By combining $\mathbf{W}_k \succeq  \mathbf{w}_k \mathbf{w}_k^H, \label{eq:Www3}$ with an additional LMI constraint, given as $\operatorname{tr}(\mathbf{W}_k)  \leq  \operatorname{tr}(\mathbf{w}_k \mathbf{w}^H_k)$, we can guarantee that Case II always holds.
We remark that $\operatorname{tr}(\mathbf{w}_k \mathbf{w}_k^H) = \operatorname{tr}(\mathbf{w}^H_k \mathbf{w}_k) =\mathbf{w}^H_k \mathbf{w}_k$. Further applying the Schur complement, ${\bf W}_k =\textbf{w}_k{\bf w}_k^H$ can be equivalently transformed into the following LMI, given as 
\begin{align}
	& \begin{bmatrix}
		\mathbf{W}_k      & \mathbf{w}_k       \\
		\mathbf{w}_k^H     & 1    
	\end{bmatrix} \succeq \mathbf{0} ,  \forall k, \operatorname{tr}(\mathbf{W}_k) - \mathbf{w}^H_k \mathbf{w}_k \leq \mathbf{0}, \forall k,
\end{align}
which completes the proof.}

\section*{Appendix B}
For $K = 1 $, we can derive that $\mathbf{h}_k^H \hat{\mathbf{W}}^\ast \mathbf{h}_k = \mathbf{h}_k^H \mathbf{W}^\ast \mathbf{h}_k$. Hence, the received SNR and the transmission rate at the user does not decrease. Besides, we have 
\begin{align}
\mathbf{W}^\ast - \hat{\mathbf{W}}^\ast  = \left(  \mathbf{W}^\ast  \right)^{\frac{1}{2}}
 \left( \mathbf{I} - \frac{  \left(\mathbf{W}^\ast\right)^{\frac{1}{2}} \mathbf{h}_k  \mathbf{h}_k^H   \left(\mathbf{W}^\ast\right)^{\frac{1}{2}}}{\mathbf{h}_k^H  \mathbf{W}^\ast \mathbf{h}_k} \right) \left(  \mathbf{W}^\ast\right)^{\frac{1}{2}} \succeq \mathbf{0},
\end{align}
indicating that the power constraint is satisfied due to ${\mathbf{W}}^\ast \succeq \hat{\mathbf{W}}^\ast $. Additionally, replacing $\mathbf{W}^\ast $ by $\hat{\mathbf{W}}^\ast $ would not decrease the transmission rate or increase the total power, showing that  $\hat{\mathbf{W}}^\ast $ is the optimum to the objective function.

Then, we discuss the case of $K > 1$ . We introduce $r = \mathbf{h}_k^H \left( \mathbf{W}_k +\sum_{i = 1,i \ne k}^{K} \mathbf{W}_i + \mathbf{R}_{\tilde{\mathbf{W}}}  + \sigma _C^2 \right)  \mathbf{h}_k -1 $ and equivalently reformulate \eqref{eq:p2all} as
\begin{subequations} \label{eq:p2appendix}
\begin{align} 
	\max_{ \{ \mathbf{W}_k, b_k\}_{k=1}^{K}, \mathbf{R}_{\tilde{\mathbf{W}}}, \lambda}~~&{\sum_{k=1}^K \log\left( 1+r  \right)  }  -  \lambda \left( {\frac{1}{\epsilon}  \text{tr}\left( \sum_{k=1}^{K} \mathbf{W}_k  + \mathbf{R}_{\tilde{\mathbf{W}}}\right) + P_0}\right) \notag \\
	&+ \sum_{k=1}^K\left(  \log b_k - b_k \left( \sum_{i = 1,i \ne k}^{K}  \mathbf{h}_k^H{{\mathbf {W}}_i}\mathbf{h}_k  + \mathbf{h}_k^H \mathbf{R}_{\tilde{\mathbf{W}}} \mathbf{h}_k + \sigma _C^2 \right) \right)     \label{eq:p2appendixa} \\
	\mathrm{s.t.} ~&  r = \mathbf{h}_k^H \left( \mathbf{W}_k +\sum_{i = 1,i \ne k}^{K} \mathbf{W}_i + \mathbf{R}_{\tilde{\mathbf{W}}}  + \sigma _C^2 \right)  \mathbf{h}_k -1 , \label{eq:p2appendixb} \\
	& \eqref{eq:p2convexb},\eqref{eq:p2convexc}, \eqref{eq:p2convexd}, \eqref{eq:p2convexe}, \eqref{eq:p2convexf} .
\end{align}
\end{subequations}

We note that with the fixed $\lambda$, problem \eqref{eq:p2appendix} is jointly convex of variables $\{ \mathbf{W}_k, b_k\}_{k=1}^{K}, \mathbf{R}_{\tilde{\mathbf{W}}}$. Thus, it can be proved that Slater's condition holds such that  strong duality holds. By introducing the Lagrange multipliers $\varpi_{k,1} \leq 0, \varpi_{k,2} \leq 0, \mu \leq 0$ and $\mathbf{\Psi}_k \succeq \mathbf{0} $, we provide the Lagrangian function of $\mathbf{W}_k$ as
\begin{align}
 \mathcal{L}(\mathbf{W}_k)  =& - \varpi_{k,1}   \mathbf{h}_k^H  \mathbf{W}_k  \mathbf{h}_k + \sum_{i = 1,i \ne k}^{K} \varpi_{i,1}   \mathbf{h}_i^H  \mathbf{W}_k   \mathbf{h}_i +  \varpi_{k,2}   \mathbf{h}_k^H  \mathbf{W}_k  \mathbf{h}_k - \sum_{i = 1,i \ne k}^{K} \varpi_{i,2}  \gamma_k \mathbf{h}_i^H  \mathbf{W}_k   \mathbf{h}_i \notag \\
& - \operatorname{tr}(\mathbf{W}_k \mathbf{\Psi}_k)+ \mu  \operatorname{tr}(\mathbf{W}_k) +  \xi ,
\end{align}
where $\xi $ represent the terms that do not involve $\mathbf{W}_k$. Then, the KKT conditions of \eqref{eq:p2appendix} is given as 
\begin{align} \label{eq:Lcondition}
\dot{\mathcal{L}}(\mathbf{W}^\ast_k) = \mathbf{0} , \mathbf{W}^\ast_k \mathbf{\Psi}_k = \mathbf{0}.
\end{align}

Then, we have $\mathbf{\Psi}^\ast_k =   \mathbf{A}_k^\ast - \varpi_{k,1}   \mathbf{h}_k^H  \mathbf{h}_k $ and
\begin{align}
	 \mathbf{A}_k^\ast = & \sum_{i = 1,i \ne k}^{K} \varpi_{i,1}   \mathbf{h}_i^H    \mathbf{h}_i +  \varpi_{k,2}   \mathbf{h}_k^H   \mathbf{h}_k - \sum_{i = 1,i \ne k}^{K} \varpi_{i,2}  \gamma_k \mathbf{h}_i^H    \mathbf{h}_i + \mu \mathbf{I}_M.
\end{align}
Nest, we discuss the rank of $ \mathbf{A}_k^\ast$ under the following cases.

1) \textbf{Case I}: $\operatorname{rank}( \mathbf{A}_k^\ast) = M$.
In this case, we have $\operatorname{rank}( \mathbf{\Psi}^\ast_k) \geq M-1$ with the inequality $\operatorname{rank}( \mathbf{X} + \mathbf{Y} ) \geq  \operatorname{rank}( \mathbf{X} ) - \operatorname{rank}( \mathbf{Y} )$~\cite{horn2012matrix}. For $ \operatorname{rank}(\mathbf{\Psi}^\ast_k ) = M $,  the first condition in \eqref{eq:Lcondition} implies $\mathbf{W}^\ast_k = \mathbf{0}$. 
For $ \operatorname{rank}(\mathbf{\Psi}^\ast_k ) = M - 1$, we have $ \operatorname{rank}( \mathbf{W}^\ast_k )= 1$.

2) \textbf{Case II}: $\operatorname{rank}( \mathbf{A}_k^\ast) = r_a < M$.
In this case, we exploit~\cite[Theorem 2]{wang2022achieving} to construct a rank-1 solution $\mathbf{W}^\ast_k $. We give $\left\lbrace \mathbf{q}_{k,i}^\ast \right\rbrace_{i=1}^{M-r_a}$to denote the columns of orthonormal basis of  $\mathbf{\Omega}_k^\ast$, which represents the nullspace of $\mathbf{A}_k^\ast$. As $\mathbf{\Psi}^\ast_k \succeq \mathbf{0}$, we have $ (\mathbf{q}_{k,i}^\ast)^H \mathbf{\Psi}^\ast_k \mathbf{q}_{k,i}^\ast = - \varpi_{k,1}  |\mathbf{h}_k^H \mathbf{q}_{k,i}^\ast |^2 \geq 0 $. Since \eqref{eq:p2appendixb} should be active at opimum indicating $\varpi_{k,1} \geq 0$, we have $\mathbf{h}_k^H \mathbf{q}_{k,i}^\ast  = 0$ and $ \mathbf{\Psi}^\ast_k  \mathbf{\Omega}_k^\ast = \mathbf{0}$. Thus, the $M - r_a$ dimensions of $\mathbf{\Psi}^\ast_k$'s null space can be represented by $\mathbf{\Omega}_k^\ast$. We further denote $\widetilde{\mathbf{\Omega}}_k^\ast$ as the null-space of $\mathbf{\Psi}^\ast_k$, we have $\operatorname{rank}(\widetilde{\mathbf{\Omega}}_k^\ast) \geq M - r_a$. Additionally, since $\operatorname{rank}( \mathbf{A}_k^\ast) = r_a $, we have $\operatorname{rank}( \mathbf{\Psi}^\ast_k) \geq r_a - 1$, which shows that $\operatorname{rank}(\widetilde{\mathbf{\Omega}}_k^\ast) \leq M - r_a + 1$. Then, it can be readily noted that $\operatorname{rank}(\widetilde{\mathbf{\Omega}}_k^\ast) =  M - r_a$ or $\operatorname{rank}(\widetilde{\mathbf{\Omega}}_k^\ast) =  M - r_a + 1 $. When $\operatorname{rank}(\widetilde{\mathbf{\Omega}}_k^\ast) =  M - r_a$ , we have $\mathbf{W}^\ast_k =  \sum_{i=1}^{M-r_a} \lambda_{k,i}^\ast \mathbf{q}_{k,i}^\ast (\mathbf{q}_{k,i}^\ast)^H$ with $\lambda_{k,i}^\ast \geq 0$. In such a case, $ \mathbf{h}_k^H \mathbf{W}_k^\ast \mathbf{h}_k = 0$, which constradicts the optimality. Hence, we conclude that $\operatorname{rank}(\widetilde{\mathbf{\Omega}}_k^\ast) =  M - r_a + 1 $. Denoting $\widetilde{\mathbf{\Omega}}_k^\ast$ as $[\mathbf{\Omega}_k^\ast, \mathbf{p}_k^\ast]$, the optimal solution $\mathbf{W}^\ast_k$ can be given as $\mathbf{W}^\ast_k =  \sum_{i=1}^{M-r_a} \lambda_{k,i}^\ast \mathbf{q}_{k,i}^\ast (\mathbf{q}_{k,i}^\ast)^H + \tilde{\lambda}^\ast_k \mathbf{p}_k^\ast (\mathbf{p}_k^\ast)^H$ with $ \tilde{\lambda}^\ast_k \geq 0$. Therefore, a rank-1 solution can be constructed as  
\begin{align}
	\hat{\mathbf{W}}_k^\ast = \mathbf{W}^\ast_k - \sum_{i=1}^{M-r_a} \lambda_{k,i}^\ast \mathbf{q}_{k,i}^\ast (\mathbf{q}_{k,i}^\ast)^H = \tilde{\lambda}^\ast_k \mathbf{p}_k^\ast (\mathbf{p}_k^\ast)^H , \hat{\mathbf{R}}^\ast_{\tilde{\mathbf{W}}} =  \mathbf{R}^\ast_{\tilde{\mathbf{W}}} +  \sum_{i=1}^{M-r_a} \lambda_{k,i}^\ast \mathbf{q}_{k,i}^\ast (\mathbf{q}_{k,i}^\ast)^H.
\end{align}

In the following, we show that the reconstructed solution, $ 	\hat{\mathbf{W}}_k^\ast $ and $\hat{\mathbf{R}}^\ast_{\tilde{\mathbf{W}}} $ satisfy the constraints. Firstly, we have 
\begin{align}
\mathbf{h}_k^H \mathbf{W}_k^\ast \mathbf{h}_k = \mathbf{h}_k^H \hat{\mathbf{W}}_k^\ast \mathbf{h}_k,    \mathbf{h}_k^H \left(\sum_{i = 1,i \ne k}^{K} \mathbf{W}^\ast_i + {\mathbf{R}}^\ast_{\tilde{\mathbf{W}}}  \right) \mathbf{h}_k =   \mathbf{h}_k^H \left(\sum_{i = 1,i \ne k}^{K} \hat{\mathbf{W}}^\ast_i + \hat{\mathbf{R}}^\ast_{\tilde{\mathbf{W}}} \right) \mathbf{h}_k.
\end{align}
Therefore, the right-hand side term in \eqref{eq:p2appendixb} and the left-hand side term in \eqref{eq:p2d} remain unchanged. 
Besides, it can be readily verified that constraints \eqref{eq:p2b} and \eqref{eq:p2c} hold, since $\mathbf{W}_k^\ast + \mathbf{R}^\ast_{\tilde{\mathbf{W}}} = \hat{\mathbf{W}}^\ast_k + \hat{\mathbf{R}}^\ast_{\tilde{\mathbf{W}}} $, which completes the proof.

\bibliographystyle{IEEEtran}
\bibliography{eebib}

\vspace{12pt}

\end{document}